\documentclass[aps,showpacs,preprint,superscriptaddress,nofootinbib]{revtex4}
\usepackage{graphicx}
\usepackage{subfigure}
\usepackage{float}
\usepackage{amsmath}
\usepackage{amsfonts}
\usepackage{color}
\usepackage{bm}
\usepackage{bbm}
\usepackage{txfonts}
\usepackage{array}
\usepackage{amssymb}

\begin{document}

\title{Momentum spirals in multiphoton pair production revisited}
\author{Li-Na Hu}
\affiliation{Key Laboratory of Beam Technology of the Ministry of Education, and College of Nuclear Science and Technology, Beijing Normal University, Beijing 100875, China}
\author{Orkash Amat}
\affiliation{Key Laboratory of Beam Technology of the Ministry of Education, and College of Nuclear Science and Technology, Beijing Normal University, Beijing 100875, China}
\author{Li Wang}
\affiliation{Institute of Radiation Technology, Beijing Academy of Science and Technology, Beijing 100875, China}
\author{Adiljan Sawut}
\affiliation{Key Laboratory for GeoMechanics and Deep Underground Engineering, China University of Mining and Technology, Beijing 100083, China}
\author{Hong-Hao Fan}
\affiliation{Key Laboratory of Beam Technology of the Ministry of Education, and College of Nuclear Science and Technology, Beijing Normal University, Beijing 100875, China}
\author{B. S. Xie \footnote{bsxie@bnu.edu.cn}}
\affiliation{Key Laboratory of Beam Technology of the Ministry of Education, and College of Nuclear Science and Technology, Beijing Normal University, Beijing 100875, China}
\affiliation{Institute of Radiation Technology, Beijing Academy of Science and Technology, Beijing 100875, China}
\date{\today}

\begin{abstract}
Spirals in multiphoton pair production are revisited by two counter-rotating fields with time delay for different cycles in pulse. Novel findings include that for subcycle fields, the remarkable spiral structure in the momentum spectrum can be still caused by a large time delay compared to the previous study for supercycle case where it is easier to be generated by a small time delay. And also there exist a range of critical polarization values for the spirals appearance corresponding to the different cycle number. The relative phase difference between two fields causes not only severe symmetry breaking of the momentum spectra pattern and spiral, but also a significant change for the shape and the number of spiral arm. Upon the number density, it is found a more sensitive to the cycle number, in particularly, it is enhanced by more than one order of magnitude for small cycle pulse, while it is increased about few times when the time delay is small. These results provide an abundant theoretical testbed for the possible experimental observation on the multiphoton pair production in future. Meanwhile, it is applicable to regard the particles momentum signatures as a new probing to the laser field information with it from the vacuum.
\end{abstract}
\pacs{12.20.Ds, 11.15.Tk}
\maketitle

\section{Introduction}

In past decades, the researches on the electron-positron ($e^{-}e^{+}$) pair production from vacuum in strong background fields have attracted many people interest and many works have been performed theoretically \cite{Sauter:1931zz,Heisenberg:1935qt,DiPiazza:2011tq,Fedotov:2022,Schwinger:1951nm,Xie:2017,Dunne:2008kc}, while the Schwinger critical field strength $E_{\mathrm{cr}}=m^2c^{3} / e \hbar \approx 1.3 \times 10 ^{16}~\rm {\mathrm{V}/\mathrm{cm}}$ (where $m$ and $-e$ is the electron mass and charge) is still a few orders higher than present laser field and also ones of planned laser facilities such as the Extreme Light Infrastructure \cite{ELI}, the Exawatt Center for Extreme Light Studies, and the x-ray free electron laser \cite{XCELS}. In $1997$, however, an impressive E-$144$ experiment has been performed at Stanford Linear Accelerator Center (SLAC) using $46.6\mathrm{GeV}$ electrons colliding with a laser about $10^{18}\mathrm{W/cm^2}$ \cite{Burke:1997ew}, by which it is observed the production of $4-5$ pairs of $e^{-}e^{+}$. Intrigued by this multiphoton pair production experiment and also with the rapid development of high-intensity laser technology \cite{XFEL,Ringwald:2001ib,Corkum:2007,Hern:2013}, multiphoton pair creation mechanism provides a more experimental chances in future. Some new important development that include the ponderomotive force effect \cite{Kohlfurst:2017hbd}, node structures \cite{Li:2015Non} and effective mass signatures \cite{Kohlfurst:2013ura} are revealed.

Recently, spirals have attracted people more and more attention, for example, the spirals structures of photoelectron momentum distributions are identified in the multiphoton ionization under two counter-rotating circularly polarization (CP) fields \cite{Ngoko:2015Ele,Pengel:2017Elec}. The momentum spirals in photodetachment from the H$^{-}$ driven by pairs of counter-rotating CP pulses are revealed in Ref. \cite{Majczak:2022}, meanwhile, it is reported the formation of electron vortices \footnote{Note that the roots of vortices can be traced back to the Ref. \cite{Dirac:1931}, in which Dirac showed that in three-dimensional space the vortex motion is described by lines where the complex amplitude vanishes and that the circulation of the probability current around the contour containing such a line does not disappear and is quantized. It is known that vortices are different from spirals, in our work, we focus on the spiral structures in the momentum distributions of created particles.} in the same photodetachment driven by single CP pulse or pairs of co-rotating CP pulses. In fact, spirals have been widely investigated in atomic and molecular ionization \cite{Macek2009,K2016,L2020}, nonlinear optics \cite{Harris1994}, type-II superconductors \cite{Blatter1994}, plasmas physics \cite{Uby1995,Shukla2009}, atomic condensates \cite{Madison2000}, and so on. Interestingly, our previous studies showed that there exist also the significant spiral structures in multiphoton pair production \cite{Li:2017qwd,Li:2019hzi}. The spirals formed with an even number of spiral arms in one-color two counter-rotating CP laser fields with time delay \footnote{About the time delay, some researches have been performed for the Compton radiation spectra \cite{Krajewska:2014} and the Breit-Wheeler pair production \cite{Jansen:2017}.} is reported in Ref. \cite{Li:2017qwd}. Then the spirals constituted by an odd number of spiral arms in two-color counter-rotating elliptically polarization fields with time delay are also discovered \cite{Li:2019hzi}. These primary studies have indicated that the spirals in multiphoton pair production are sensitive to the field parameters.

On the other hand, it should be noticed that the previous researches have been worked within some limited range, for instance, either the cycle or the time delay between two fields is fixed, etc. However, for various cycles and time delays, is there still a spiral in multiphoton pair creation? How about the spiral changes when the relative phase is introduced between two fields? Since the momentum pattern and spiral is very sensitive to the ellipticity of polarized fields, what is the range of ellipticity to observe spirals effectively?

For the purpose to clarify points mentioned above, therefore, in this paper, we shall revisit the spirals in multiphoton pair production in two counter-rotating fields with time delay by using the Dirac-Heisenberg-Wigner (DHW) formalism. The studying is focusing on the effects of time delay and the number of cycles in pulse on the momentum spiral and the number density of created particles, in two typical cases of relative carrier envelope phase as $0$ and $\pi/2$, respectively. Without losing generality, we shall consider four different cases of time delay and three different cycles of supercycle, subcycle and cycle between them. It is found that there is still an obvious spiral structure in the momentum spectrum even if in the case of subcycle. Some novel features and interesting phenomena for the spiral would be revealed.

We consider the following spatially homogenous and time-varying electric field model that is composed of two counter-rotating fields with time delay \cite{Li:2017qwd,Titov:2018bgy},
\begin{eqnarray}\label{model}
\mathbf{E}(t)=\mathbf{E}_{1}(t)+\mathbf{E}_{2}(t),
\end{eqnarray}
with
\begin{eqnarray}
\mathbf{E}_{1,2}(t)=f_{1,2}(t)\mathbf{g}_{1,2}(t),
\end{eqnarray}
where
\begin{eqnarray}
f_{1,2}(t)=\frac{E_{1,2}}{\cosh(\frac{t\pm T}{\tau})} \, ,
\quad  \mathbf{g}_{1,2}(t)=[\cos(\omega(t\pm T)+\phi_{1,2}), \delta_{1,2}\sin(\omega(t\pm T)+\phi_{1,2}), 0]^\textsf{T} \,.
\end{eqnarray}
Here the sign $\textsf{T}$ denotes the transpose of the matrix, $E_{1,2}=E_0/\sqrt{1+\delta_{1,2}^2}$ are the electric field strength, $|\delta_{1,2}|=1$ denote the circular polarizations (where we define $\delta_{1}=-1$ as a right-handed CP field and $\delta_{2}=1$ as a left-handed CP field \cite{Li:2017qwd}), $\omega$ represents the field frequency, $\phi_{1,2}$ are the carrier envelope phases (the corresponding relative phase is $\Delta\phi=\phi_{2}-\phi_{1}$). And $\tau=N\pi/\omega$ is the pulse duration, where $N$ has the meaning of a number of cycles in the individual pulse. $T$ denotes the time delay parameter of two consecutive pulses at $\pm G\tau$, where $G$ is a dimensionless quantity. Accordingly, the time delay between the centers of the two consecutive pulses amounts $2T$. Since the main interest in this study is the dependence on the time delay $T$ and the number of cycles $N$ in the single pulse.

\begin{figure}[H]
\begin{center}
\includegraphics[width=\textwidth]{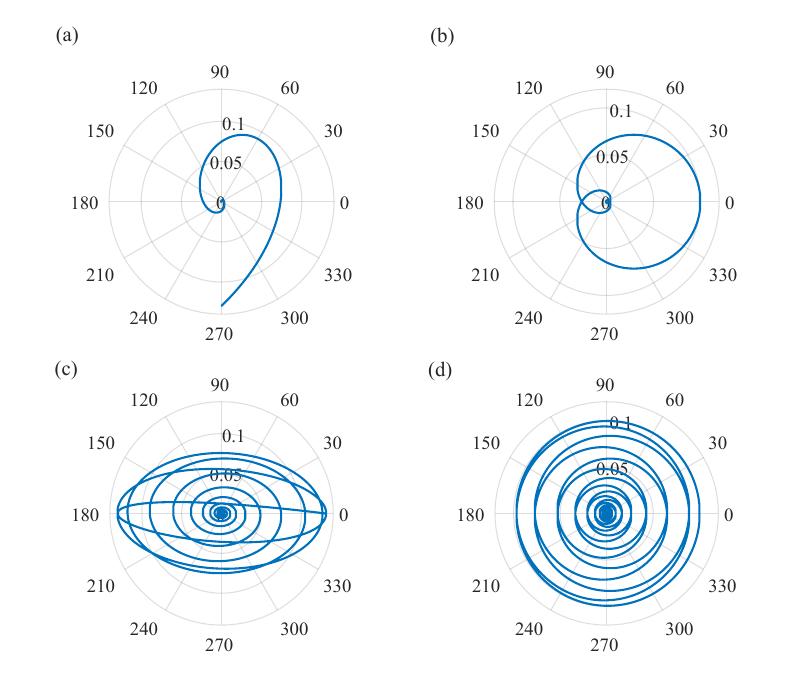}
\end{center}
\vspace{-0.01cm}
\setlength{\abovecaptionskip}{-0.6cm}
\caption{(color online). A set of typical polar diagram of the electric field Eq. (\ref{model}) for different time delays and cycles. (a) and (b) correspond to $T=\tau$ and $T=8\tau$ with $N=0.5$, respectively. (c) and (d) are $T=\tau$ and $T=8\tau$ with $N=4$. Other field parameters are $E_{1,2}=0.1\sqrt{2}E_{\mathrm{cr}}$, $\delta_{1}=-1$, $\delta_{2}=1$, $\omega=0.6$ and $\phi_{1,2}=0$.}
\label{1}
\end{figure}

A set of typical polar diagram of electric field is shown in Fig.\ref{1}. Since Eq. (\ref{model}) is consisted by two fields, where the first is a right-handed CP and the second is a left-handed CP. From Fig.~\ref{1}(a), one can see that the corresponding curve evolves clockwise from $0$ to the maximum, and then counterclockwise from the maximum to $0$. From Figs.~\ref{1}(b), (c) and (d), however, it is found that the curves begin and end at the origin of coordinates for both of fields, in which the curves evolve clockwise and counterclockwise with respect to first and second field, respectively. These properties are also reflected in the momentum spectrum of created particles. Note that the curve in Fig.~\ref{1}(a) has no self crossing for the single field of either first or second one. This is because that the $N=0.5$ and $T=\tau$ corresponds to an ultrashort pulse with subcycle, which cannot separate two fields far away so that two fields are overlapping for the first right CP and the second left CP field.

Note that, throughout this paper, we set $E_0=0.1\sqrt{2}E_\mathrm{cr}$, $\omega=0.6$ and $\phi_1=0$. And also the natural units $\hbar=c=1$ are applied and all quantities are presented in terms of the electron mass $m$. For example, the field frequency and the momentum are in unit of $m$, and the temporal scales of the electric field is in unit of $1/m$.

The paper is organized as follows. In Sec. \ref{DHW}, we briefly recall the DHW formalism to be self-contained in this work. In Sec. \ref{Momentum1}, we examine the different spiral structures and signatures with different chosen parameters of time delay and cycle number of fields when the relative carrier envelope phase is set as $0$. The case of the relative phase as $\pi/2$ is investigated in Sec. \ref{Momentum2}. In Sec. \ref{density}, the number density is presented and analyzed. Finally, the conclusion and outlook are given in Sec. \ref{conclusion}.

\section{DHW formalism}\label{DHW}

The present study is based on the DHW formalism that has been widely adopted to investigate vacuum pair production in strong background field \cite{Ababekri:2019dkl,Hebenstreit:2011wk,Kohlf2020Effect,Kohlfurst:2017git,Li:2021vjf,Hebenstreit:2011pm,Kohlfurst:2015zxi}. Since the detailed derivation of the DHW formalism has been performed in Refs. \cite{Olugh:2018seh,Li:2015cea,Olugh:2019nej}, here we only present the key points of this approach.

We start from the gauge-covariant density operator of two Dirac field operators in the Heisenberg picture,
\begin{equation}\label{DensityOperator}
 \hat {\mathcal C}_{\alpha \beta} \left(r , s \right) = \mathcal U \left(A,r,s
\right) \ \left[ \bar \psi_\beta \left( r - s/2 \right), \psi_\alpha \left( r +
s/2 \right) \right],
\end{equation}
where $r$ denotes the center-of-mass coordinate and $s$ the relative coordinate. The Wilson line factor
\begin{equation}\label{Wilson line factor}
\mathcal U \left(A,r,s \right) = \exp\left(\mathrm{i}\ e\ s\int_{-1/2}^{1/2} d
\xi \ A \left(r+ \xi s \right)  \right),
\end{equation}
is used to guarantee the density operator gauge invariant, and it is related to the elementary charge $e$ and the background gauge field $A$.

It is known that the important quantity of the DHW approach is the covariant Wigner operator, which could be defined as the Fourier transform of Eq. \eqref{DensityOperator} with respect to the relative coordinate $s$, i.e.,
\begin{equation}\label{WignerOperator}
\hat{\mathcal W}_{\alpha \beta} \left( r , p \right) = \frac{1}{2} \int d^4 s \
\mathrm{e}^{\mathrm{i} ps} \  \hat{\mathcal C}_{\alpha \beta} \left( r , s
\right).
\end{equation}
By taking the vacuum expectation value of Eq. \eqref{WignerOperator}, we can obtain the covariant Wigner function
\begin{equation}\label{Wigner function}
 \mathbbm{W} \left( r,p \right) = \langle \Phi \vert \hat{\mathcal W} \left( r,p
\right) \vert \Phi \rangle.
\end{equation}
Because of the fact that the Wigner function is in the Dirac algebra, it can be decomposed into $16$ covariant Wigner coefficients
\begin{equation}\label{decomposed}
\mathbbm{W} = \frac{1}{4} \left( \mathbbm{1} \mathbbm{S} + \textrm{i} \gamma_5
\mathbbm{P} + \gamma^{\mu} \mathbbm{V}_{\mu} + \gamma^{\mu} \gamma_5
\mathbbm{A}_{\mu} + \sigma^{\mu \nu} \mathbbm{T}_{\mu \nu} \right),\
\end{equation}
where $\mathbbm{S}$, $\mathbbm{P}$, $\mathbbm{V}_{\mu}$, $\mathbbm{A}_{\mu}$ and $\mathbbm{T}_{\mu \nu}$ denote scalar, pseudoscalar, vector, axial vector and tensor, respectively. According to the Refs. \cite{Hebenstreit:2011pm,Kohlfurst:2015zxi,Bialynicki-Birula:1991jwl,Hebenstreit:2010vz}, the equations of motion for the Wigner function can be written as
\begin{equation}
D_{t}\mathbbm{W} = -\frac{1}{2}\mathbf{D}_{\mathbf{x}}[\gamma^{0}\bm{\gamma},\mathbbm{W}]
+\mathrm{i}m[\gamma^{0},\mathbbm{W}]-\mathrm{i}\mathbf{P}\{\gamma^{0}\bm{\gamma},\mathbbm{W}\},
\label{motion}
\end{equation}
here $D_{t}$, $\mathbf{D}_{\mathbf{x}}$ and $\mathbf{P}$ represent the pseudodifferential operators.

Inserting Eq. \eqref{decomposed} into Eq. \eqref{motion}, we can get a set of partial differential equations (PDEs) for the $16$ Wigner components. For the spatially uniform and time-dependent electric field Eq. (\ref{model}), by applying the method of characteristics \cite{Hebenstreit:2011wk,Bialynicki-Birula:1991jwl,Hebenstreit:2010vz,Blinne:2013via,Blinne:2015zpa} and replacing the kinetic momentum ${\mathbf p}$ with the canonical momentum ${\mathbf q}$ via $ {\mathbf q} - e {\mathbf A} (t)$, the PDEs for the $16$ Wigner components can be simplified to the ordinary differential equations (ODEs) for only $10$ Wigner components. And the corresponding Wigner coefficients are
\begin{equation}
{\mathbbm w} = ( {\mathbbm s},{\mathbbm v}_i,{\mathbbm a}_i,{\mathbbm t}_i)
\, , \quad  {\mathbbm t}_i := {\mathbbm t}_{0i} - {\mathbbm t}_{i0}  \, .
\end{equation}
For the specific derivation of these $10$ equations, we refer the reader to Refs. \cite{Hebenstreit:2011pm,Kohlfurst:2015zxi,Olugh:2018seh,Olugh:2019nej}. In order to perform calculations, the vacuum initial conditions are given \cite{Li:2015cea,Olugh:2018seh} by
\begin{equation}
{\mathbbm s}_{vac} = \frac{-2m}{\sqrt{{\mathbf p}^2+m^2}} \, ,
\quad  {\mathbbm v}_{i,vac} = \frac{-2{ p_i} }{\sqrt{{\mathbf p}^2+m^2}} \, .
\end{equation}
The single-particle momentum distribution function is defined as
\begin{equation}
f({\mathbf q},t) = \frac 1 {2 \Omega(\mathbf{q},t)} (\varepsilon - \varepsilon_{vac} ),
\end{equation}
where $\Omega(\mathbf{q},t)= \sqrt{m^2+{\mathbf p}^2(t)}=\sqrt{m^{2}+(\mathbf{q}-e\mathbf{A}(t))^{2}}$ denotes the total energy of particles, $\varepsilon = m {\mathbbm s} + p_i {\mathbbm v}_i$ represents the phase space energy density. To precisely calculate the distribution function $f({\mathbf q},t)$, it is necessary to introduce an auxiliary three-dimensional vector \cite{Blinne:2013via,Blinne:2015zpa}
\begin{equation}
\mathbf{v} (\mathbf{q},t) : = {\mathbbm v}_i (\mathbf{p}(t),t) -
(1-f({\mathbf q},t))  {\mathbbm v}_{i,vac} (\mathbf{p}(t),t) \, .
\end{equation}
Therefore, we can obtain the single-particle momentum distribution function $f({\mathbf q},t)$ by solving the following ODEs
\begin{equation}
\begin{array}{l}
\dot{f}=\frac{e\mathbf{E}\cdot \mathbf{v}}{2\Omega},\\
\dot{\mathbf{v}}=\frac{2}{\Omega^{3}}[(e\mathbf{E}\cdot \mathbf{p})\mathbf{p}-e\mathbf{E}\Omega^{2}](f-1)-\frac{(e\mathbf{E}\cdot \mathbf{v})\mathbf{p}}{\Omega^{2}}-2\mathbf{p}\times \mathbbm{a}-2m\mathbbm{t},\\
\dot{\mathbbm{a}}=-2\mathbf{p}\times \mathbf{v},\\
\dot{\mathbbm{t}}=\frac{2}{m}[m^{2}\mathbf{v}+(\mathbf{p}\cdot \mathbf{v})\mathbf{p}],
\end{array}\label{eq3}
\end{equation}
with the initial conditions $f(\mathbf{q},-\infty)=0$, $\mathbf{v}(\mathbf{q},-\infty)=\mathbbm{a}(\mathbf{q},-\infty)=\mathbbm{t}(\mathbf{q},-\infty)=0$. Here the dot represents a total time derivative, $\mathbf{v}$, $\mathbbm{a}$ and $\mathbbm{t}$ are the three-dimensional vectors corresponding to Wigner components, and their physical sense is as follows, $\mathbf{v}$ denotes current density, $\mathbbm{a}$ is spin density, and $\mathbbm{t}$ is magnetic moment density \cite{Kohlf2019S}. $\mathbf{E}$ is our electric field of Eq. (\ref{model}), $\Omega=\sqrt{m^2+{\mathbf p}^2}=\sqrt{m^{2}+(\mathbf{q}-e\mathbf{A}(t))^{2}}$ is the total energy of particles, where ${\mathbf p}$ represents kinetic momentum and $\mathbf{q}$ denotes canonical momentum, and $e$ is the charge of particle, i.e., $|e|$ and $-|e|$ for positron and electron, respectively \footnote{In fact, the sign of $e$ does not affect scientific results, and the $\pm$ of $e$ corresponds to the momentum spectrum of created positron/electron. Equivalently, if the sign of $e$ changes meanwhile the electric field reverses its sign, the momentum spectrum of created particles is consistent with each other. In present work we choose the positive charge.}. $\mathbf{A}(t)$ is the vector potential of the external field. By the way, in solving Eq. \eqref{eq3} above, we employ fortran software by the Runge-Kutta $4$th order with a fixed time step $0.0005$. And the number of lattice points of momentums $q_{x}$, $q_{y}$ are set as $N_{q_{x}}=N_{q_{y}}=800$, refer to \cite{Kohlf2020Effect,Kohlfurst:2015zxi}.

Moreover, the number density of created pairs can also be obtained by integrating the distribution function $f(\mathbf{q},t)$ over full momenta at $t\rightarrow+\infty$, i.e.,
\begin{equation}\label{14}
  n = \lim_{t\to +\infty}\int\frac{d^{3}q}{(2\pi)^ 3}f(\mathbf{q},t) \, .
\end{equation}

\section{Spirals for fields with relative phase $\Delta\phi=0$}\label{Momentum1}

In this section, we study the effects of time delay with different cycles in pulse on the momentum spirals in multiphoton pair production by two counter-rotating fields with relative carrier envelope phase $\Delta\phi=\phi_2-\phi_1=0$.

\subsection{$N=4$}

\begin{figure}[H]
\vspace{-1em}
\begin{center}
\includegraphics[width=\textwidth]{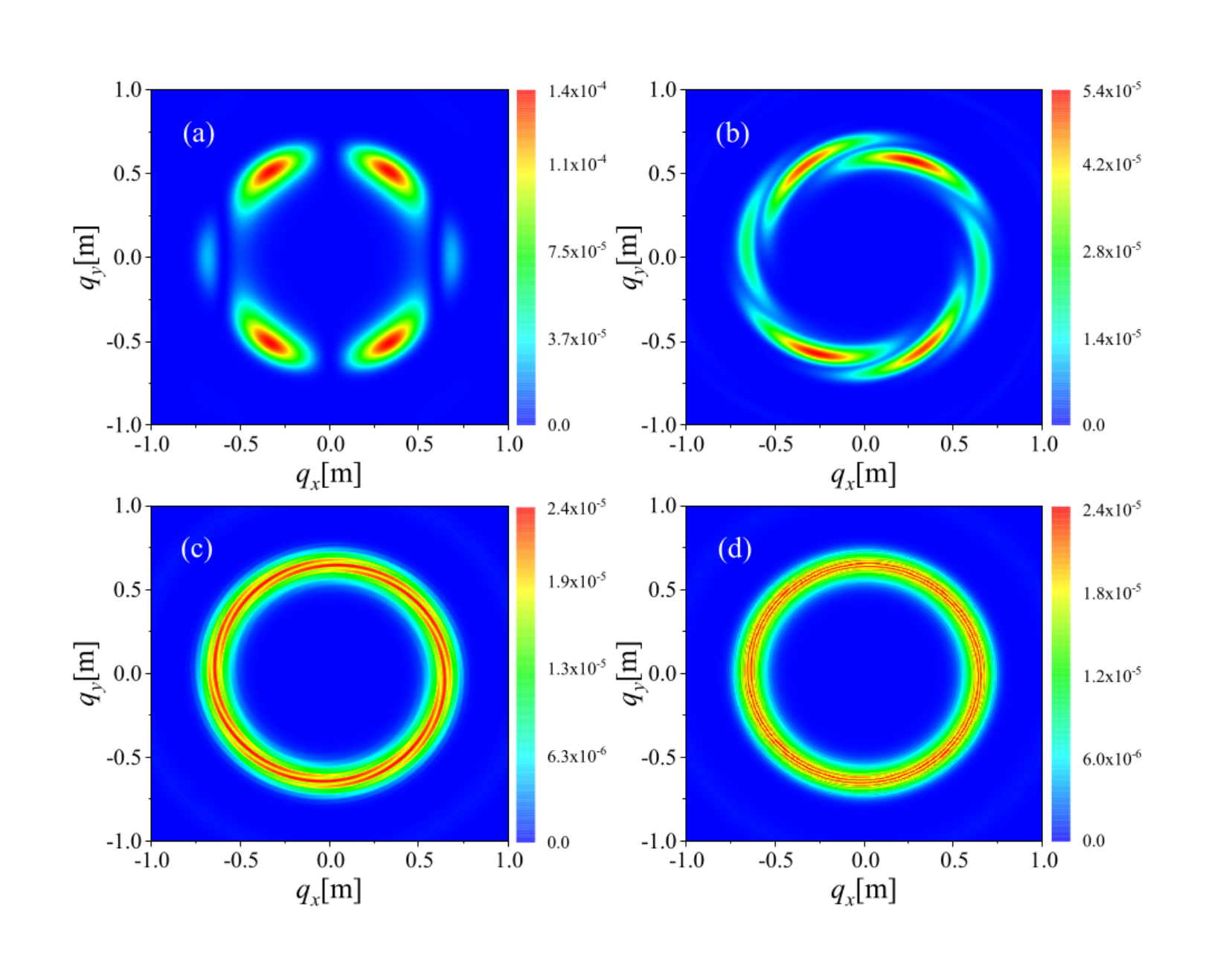}
\end{center}
\setlength{\abovecaptionskip}{-1.6cm}
\caption{(color online). Momentum spectra of created particles in the polarization plane (where $q_z=0$) for $N=4$ with different time delay parameters. From (a) to (d), the corresponding time delays are $T=0, \tau, 4\tau, 8\tau$, respectively. Other electric field parameters are $E_{1,2}=0.1\sqrt{2}E_{\mathrm{cr}}$, $\delta_{1}=-1$, $\delta_{2}=1$, $\omega=0.6$, and $\phi_{1,2}=0$.}
\label{2}
\end{figure}

We know that the time delay is fixed in previous study \cite{Li:2019hzi}, but now we explore how the momentum spectrum changes in the case of varying time delay. Before presenting our findings in detail, the symmetry of the momentum spectrum in the case $T=0$ is briefly discussed. When $N=4$, the effects of $T$ on the momentum spectra in the polarization plane for two counter-rotating fields are shown in Fig.~\ref{2}. For $T=0$, one can see that the momentum spectrum presents four bright curved moon-shaped structures and has a good axisymmetry in the $q_{x}$ and $q_{y}$ directions, see Fig.~\ref{2}(a). Since the momentum distribution is mainly related to the total energy of particle $\Omega(\mathbf{q},t)=\sqrt{m^{2}+(\mathbf{q}-e\mathbf{A}(t))^{2}}=\sqrt{m^{2}+(q_{x}-eA_{x}(t))^{2}+(q_{y}-eA_{y}(t))^{2}}$, and only in the case of $T=0$, $A_{x}(t)$ is an odd function with respect to $t$ while $A_{y}(t)=0$. Therefore, under time reversal, the time $t$ and the momentums $q_{x}$ and $q_{y}$ change sign, $\Omega(\mathbf{q},t)$ still stays invariant, which ensures a good axisymmetry of the momentum spectrum. Once $T=\tau$, the symmetry in the $q_{x}$ direction still exists, but the symmetry in the $q_{y}$ direction gradually disappears.

In the case of varying $T$, our findings include that as time delay increases to $T=\tau$, the four curved moon-shaped structures in Fig.~\ref{2}(a) are gradually elongated and rotated, which eventually leads to the generation of spiral structure in the momentum spectrum, see Fig.~\ref{2}(b). Importantly, it is found that the spiral is consisted of six spiral arms. With the time delay increasing to $T=4\tau$ and $T=8\tau$, however, we observed that the momentum spectra exhibit the signature of eight arms of spiral pattern, see Figs.~\ref{2}(c) and (d). And compared to the case of $T=\tau$, the spiral arms become longer and slender, resulting in the appearance of a more pronounced spiral structure. This phenomenon may be understood qualitatively from the evolution of the electric field in Figs.~\ref{1}(c) and (d). One can see that the evolution curve for $T=8\tau$ in Fig.~\ref{1}(d) presents a wider and more uniform distribution than that of $T=\tau$, which leads to a more pronounced spiral structure in the corresponding momentum spectrum. In particular, for $T=8\tau$, the spiral pattern becomes almost a quasi-Rasmey interference fringe consisting of many concentric rings.

Upon the particle spin effect on the pair production, there has been some studies \cite{Blinne:2015zpa,Strobel:2014tha,Kohlf2019S,Li:2019}. For the single field, there exists some difference for the effect of spin-up or spin-down of particles on the pair creation \cite{Blinne:2015zpa,Strobel:2014tha,Kohlf2019S}. Under the two counter-rotating fields with time delay, however, as what shown in Ref. \cite{Li:2019hzi,Li:2019} as that the effect of particle spin-up and spin-down on the spiral structures in the momentum spectrum of created particles can be ignored since the invariance of the combinational symmetry by opposite helicity of two CP field and the particle spin. Therefore, in present work, we do not need to consider it.

In order to clearly understand the spiral structures in the momentum spectra described above, we employ the Wentzel-Kramers-Brillouin (WKB)-like approximation method \cite{Blinne:2015zpa,Strobel:2014tha} to make some semiquantitative understandings on obtained numerical results. It is known that $e^{-}e^{+}$ pairs are primarily created at the maximum of the electric field, i.e., at $t=-T$ and $t=T$ for the electric field Eq. (\ref{model}), and the creation process is dominated by the two pairs of turning points near $t=-T$ and $t=T$. According to WKB-like approximation \cite{Li:2017qwd,Li:2019hzi,Blinne:2015zpa,Strobel:2014tha,Akkermans:2011yn}, for a certain $\mathbf{q}$, the amplitude of pair creation for the first field in our model can be expressed as $A_1=\exp[-iK_s(\mathbf{q},t_1^+)]$ and correspondingly, the second one can be written as $A_2=\exp[-iK_s(\mathbf{q},t_2^+)]$, where $t_1$ and $t_2$ represent the turning points near $t=-T$ and $t=T$. Therefore, one can obtain the momentum distribution function
\begin{eqnarray}\label{D1}
f(\mathbf{q})&=&\sum_{s=\pm}|A_1+A_2|^2 \nonumber \\
&=&\sum_{s=\pm}\Big|e^{-iK_s(\mathbf{q},t_1^+)}+e^{-iK_s(\mathbf{q},t_2^+)}\Big|^2,
\end{eqnarray}
where $K_s(\mathbf{q},t)=K_0(\mathbf{q},t)-sK_{xy}(\mathbf{q},t)$, $K_0(\mathbf{q},t)=2\int_{-\infty}^t\Omega(\mathbf{q},t')dt'$,
$K_{xy}(\mathbf{q},t)= \epsilon_\perp\int_{-\infty}^t\frac{\dot{p}_x(t')p_y(t')-\dot{p}_y(t')p_x(t')}{\Omega(\mathbf{q},t')[p_x^2(t')+p_y^2(t')]}dt'$,
$s=\pm1$ represents the electron spin, $s=0$ denotes the scalar particle, $\Omega(\mathbf{q},t)=\sqrt{m^2+[\mathbf{q}-e\mathbf{A}(t)]^2}$ and $\epsilon_\perp=\sqrt{m^2+q_z^2}$. For a larger time delay $T$, the amplitude of pair production for the second field can be rewritten as $A_2=\exp[i\theta_s(\mathbf{q})]A_1$, where $\theta_s(\mathbf{q})=\mathrm{Re}[K_s(\mathbf{q},t_2^+)
-K_s(\mathbf{q},t_1^+)]$ denotes a phase accumulated factor between the two pulses \cite{Li:2017qwd,Li:2019hzi}. From these, the Eq. \eqref{D1} becomes
\begin{eqnarray}\label{D2}
f(\mathbf{q})&=&\sum_{s=\pm}|A_1+e^{i\theta_s(\mathbf{q})}A_1|^2 \nonumber \\
&=&\sum_{s=\pm}2(1+\cos[\theta_s(\mathbf{q})])e^{-2\vartheta_s(\mathbf{q},t_1^+)} \nonumber \\
&\propto&\big\{1+\cos[\theta_0(\mathbf{q})]\big\}e^{-2\vartheta_0(\mathbf{q},t_1^+)}.
\end{eqnarray}
Note that for a large $T$, since both the electric field and the vector potential between $t=-T$ and $t=T$ are very small, the $\theta_0(\mathbf{q})$ can be expressed as $\theta_0(\mathbf{q})\approx4\sqrt{m^2+\mathbf{q}^2}T$.

Actually, it is more convenient to understand the variation in the number and the shape of spiral arms in the spherical coordinates ($q,\theta,\varphi$). Since the amplitude of pair creation in spherical coordinates is able to provide a phase factor which may well reveal the rotation properties of a CP field. Similar to the Refs. \cite{Li:2017qwd,Li:2019hzi}, the amplitude $A_1$ in spherical coordinates can be written as $A_1\approx\exp(i\ell\delta_1\varphi)A_0(q,\theta,\varphi)$, where $\ell$ is the the number of photons absorbed in the multiphoton pair production process and $\varphi$ denotes the azimuthal angle, and the amplitude $A_2$ can be expressed as $A_2\approx\exp(i\ell\delta_2\varphi)\exp[i\theta_0(q,\theta,\varphi)]A_0(q,\theta,\varphi)$. Finally, combining Eqs. (\ref{D1}) and (\ref{D2}), we can obtain the momentum distribution function in the polarization plane (where the polar angle $\theta=\pi/2$, i.e., $q_z=0$)
\begin{equation}\label{D3}
f(q,\varphi)\propto\big\{1+\cos[\theta_0(q,\varphi)+(\delta_2-\delta_1)\ell\varphi]\big\}|A_0(q,\varphi)|^2,
\end{equation}
here $\theta_0(q,\varphi)\approx4\sqrt{m^2+q^2}T$ for a large $T$. It is known from Eq. (\ref{D3}) that the number of spiral arms is associated with
\begin{equation}\label{D4}
q_{k'}^\mathrm{max}(\varphi)=\sqrt{\Big[\frac{2k'\pi-(\delta_2-\delta_1)\ell\varphi}{4T}\Big]^2-m^2},
\end{equation}
where $k'$ is an integer. Furthermore, we can know from Eq. (\ref{D4}) that the spiral arms number is primarily determined by $|(\delta_{2}-\delta_{1})\ell|$, which will be illustrated in the following numerical results.

For example, when time delays are $T=4\tau$ and $T=8\tau$ in Figs.~\ref{2}(c) and (d), the helicities of the two counter-rotating CP fields are $\delta_{1}=-1$ and $\delta_{2}=1$ as well as the frequencies of those are $\omega=0.6$. According to the energy conservation equation $\ell\omega=2\sqrt{q^2+m_{*}^2}$ with the effective mass $m_{*}$, we know that the pair production is related to $4$-photon process, i.e., $\ell=4$. Therefore, one can obtain $|(\delta_{2}-\delta_{1})\ell|=8$, which indicates that the spiral pattern is composed of $8$ spiral arms. It has a good agreement with our numerical results.

In addition, the change in the shape of the spiral arm with increasing time delay can also be understood.
According Eq. (\ref{D4}), we can obtain $\varphi(q)=\big(2k'\pi-4\sqrt{q^2+m^2}T\big)/(\delta_2-\delta_1)\ell$. The absolute value of the derivative for the above equation with respect to $q$ can be written as
\begin{equation}\label{D5}
|\mathrm{d}\varphi(q)/\mathrm{d}q|=|4T/(\delta_2-\delta_1)\ell|\cdot (q/\sqrt{q^2+m^2}).
\end{equation}
One can see from Eq. (\ref{D5}) that the increase of $\varphi$ with $q$ changes more quickly for a large $T$ than a small one. It indicates that the larger the time delay, the faster the spiral structure rotates, which causes the spiral arms to become thinner, longer and tighter. These results are consistent with the variation of spirals in the momentum spectrum shown in Figs.~\ref{2}(c) and (d).

\subsection{$N=2$}

\begin{figure}[H]
\begin{center}
\includegraphics[width=\textwidth]{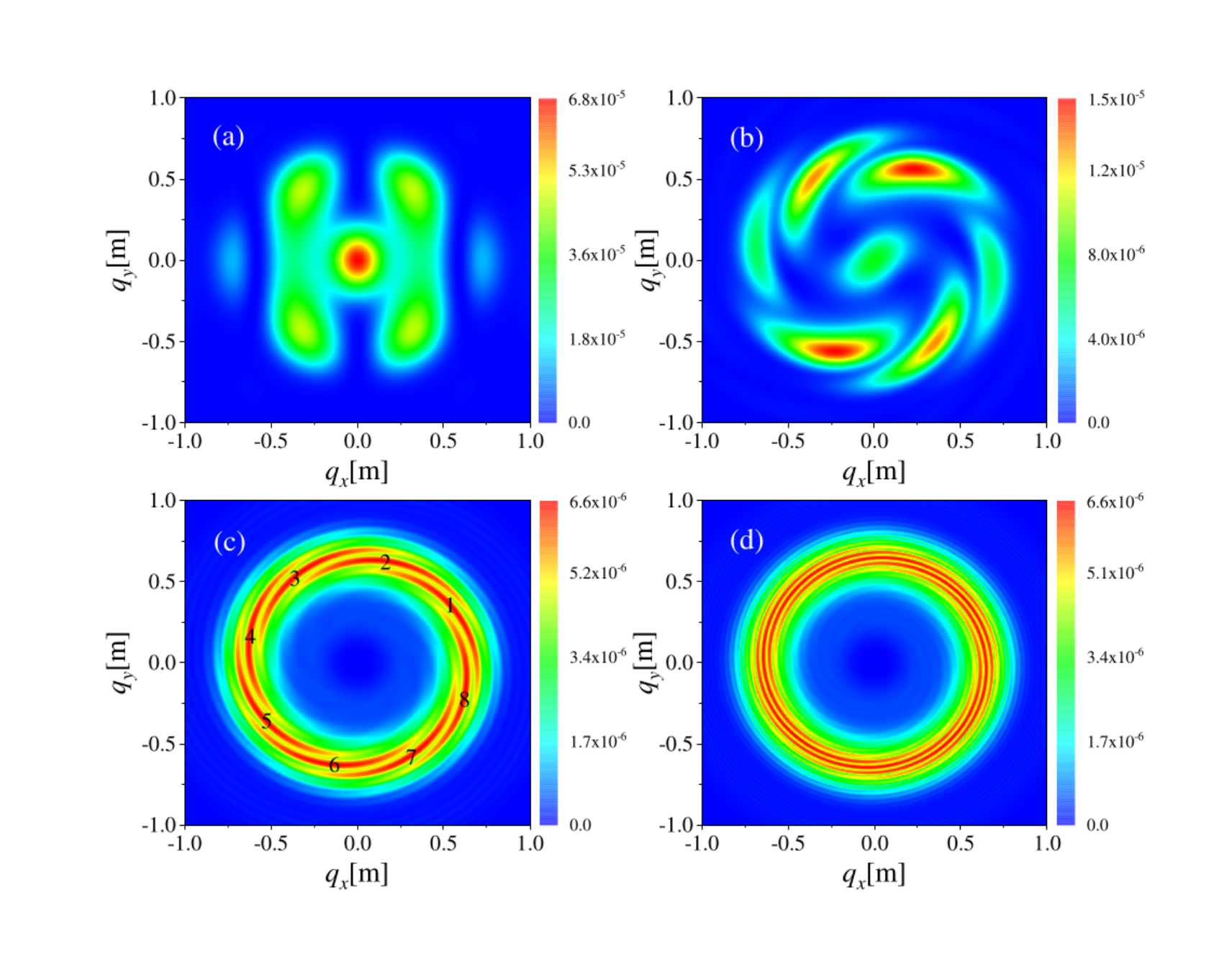}
\end{center}
\setlength{\abovecaptionskip}{-1.6cm}
\caption{(color online). Momentum spectra of created particles in the polarization plane (where $q_z=0$) for $N=2$ with different time delay parameters. From (a) to (d), the corresponding time delays are $T=0, \tau, 4\tau, 8\tau$, respectively. Other electric field parameters are the same as in Fig.~\ref{2}.}
\label{3}
\end{figure}

In our previous work \cite{Li:2017qwd}, the effect of a relatively large time delay on the momentum spectrum with relative phase $\Delta\phi=\pi/2$ is considered. Here we add some details, one is that time delay is relatively small, the other is relative phase $\Delta\phi=0$. We shows the effects of $T$ on the momentum distributions for $N=2$ in Fig.~\ref{3}. For $T=0$, the result is almost the same as in Fig.~\ref{2}(a) of Ref.\cite{Li:2017qwd}, in addition, it is similar to that of $N=4$, i.e., the momentum distribution still exists a good axisymmetry in the $q_{x}$ and $q_{y}$ directions. While as time delay increases to $T=\tau$, we found some results include that the axisymmetry is destroyed, and the H-shaped momentum distribution that is the strongest near the center in Fig.~\ref{3}(a) gradually expands outward and twists simultaneously, which eventually can cause the generation of spiral structures, see Fig.~\ref{3}(b). Meanwhile, the spiral pattern presents an obvious rotational symmetry. The reason is that even though there is a time delay between the two counter-rotating fields, under time reversal, the momentums $q_{x}$ and $q_{y}$ change sign, the total energy of particle $\Omega(\mathbf{q},t)$ remains almost invariant. Therefore, there is a pronounced rotational symmetry in the spiral structure, and the larger the time delay, the better the rotational symmetry. Moreover, we also found that the spiral is consisted of six inhomogeneous spiral arms. Since when $T$ is small, the two fields are not yet completely separated and there is an overlap between them, so there is a remarkable interference effect between them, which leads to the generation of the inhomogeneous spiral structure in Fig.~\ref{3}(b).

As time delay further increases to $T=4\tau$ and $T=8\tau$, compared to the case of $T=\tau$, we observed that the number of spiral arms changes from $6$ to $8$, and the distribution of the arms changes from inhomogeneity to homogeneity, as shown in Figs.~\ref{3}(c) and (d). Where about the understanding of the number of spiral arms is similar to the Fig.~\ref{2}. The change in homogeneity is due to the fact that when $T$ is large, the two fields are completely separated, so the interference effect between them is significantly reduced, which leads to a relatively uniform distribution of the spiral arms. In addition, according to Eq. (\ref{D4}), we can determine the position of the eight spiral arms in the momentum spectrum. For instance, in the case of $T=4\tau$, the estimated results of the maximum value positions for eight spiral arms are shown in Table~\ref{Table 1}. Compared with the numerical results in Fig.~\ref{3}(c), we found a good agreement between them, and the errors lie within at about $2\%\sim6.5\%$.

\begin{table}[H]
\caption{Comparison of $q_{x}$ ($q_{y}$) and $q_{x}^\mathrm{est}$ ($q_{y}^\mathrm{est}$) corresponding to the maximum value of the $8$ spiral arms in Fig.~\ref{3}(c), where $q_{x}$ ($q_{y}$) is the numerical result, and $q_{x}^\mathrm{est}$ ($q_{y}^\mathrm{est}$) is the estimated result of Eq. (\ref{D4}) for $k'=33,34,...,40$.}
\centering
\begin{ruledtabular}
\begin{tabular}{ccccccccc}
$i$  &$1$  &$2$  &$3$  &$4$  &$5$   &$6$   &$7$   &$8$ \\
\hline
$q_{x}$ &$0.5576$ &$0.1721$ &$-0.3259$ &$-0.6264$  &$-0.5638$  &$-0.1683$  &$0.3298$  &$0.6327$ \\
$q_{x}^\mathrm{est}$ &$0.5938$ &$0.1775$ &$-0.3429$ &$-0.6623$  &$-0.5938$  &$-0.1775$  &$0.3429$  &$0.6623$ \\
\hline
$q_{y}$ &$0.3292$ &$0.6245$ &$0.5644$ &$0.1700$  &$-0.3266$  &$-0.6295$  &$-0.5620$  &$-0.1740$ \\
$q_{y}^\mathrm{est}$ &$0.3428$ &$0.6623$ &$0.5938$ &$0.1775$  &$-0.3429$  &$-0.6623$  &$-0.5938$  &$-0.1775$ \\
\end{tabular}
\end{ruledtabular}
\label{Table 1}
\end{table}

On the other hand, it is found that the spiral patterns present an obvious difference between the cases of $N=2$ and $N=4$. For the time delay as $T=G\tau$, under the given same $G$, the spiral structure of $N=2$ is more dispersed than that of $N=4$, meanwhile, the spiral arms of $N=2$ are also shorter and thicker than those of $N=4$. These phenomena can be understood as below. Since it is known from field Eq. (\ref{model}) that the the pulse duration is $\tau=N\pi/\omega$, so that the time delay is $T=GN\pi/\omega$. It leads to the fact that when $G$ is fixed, the smaller $N$ is, the smaller the corresponding $T$ is. Then according to Eq. (\ref{D5}), we can see that for small $T$, $\varphi$ varies slowly with increasing $q$, which means that the spiral structure with $N=2$ rotates slower than that one with $N=4$. This eventually results in the spiral structure for $N=2$ being more dispersed than that of $N=4$, and the spiral arms of $N=2$ are shorter and thicker than that of $N=4$.

\subsection{$N=1$}

\begin{figure}[H]
\begin{center}
\includegraphics[width=\textwidth]{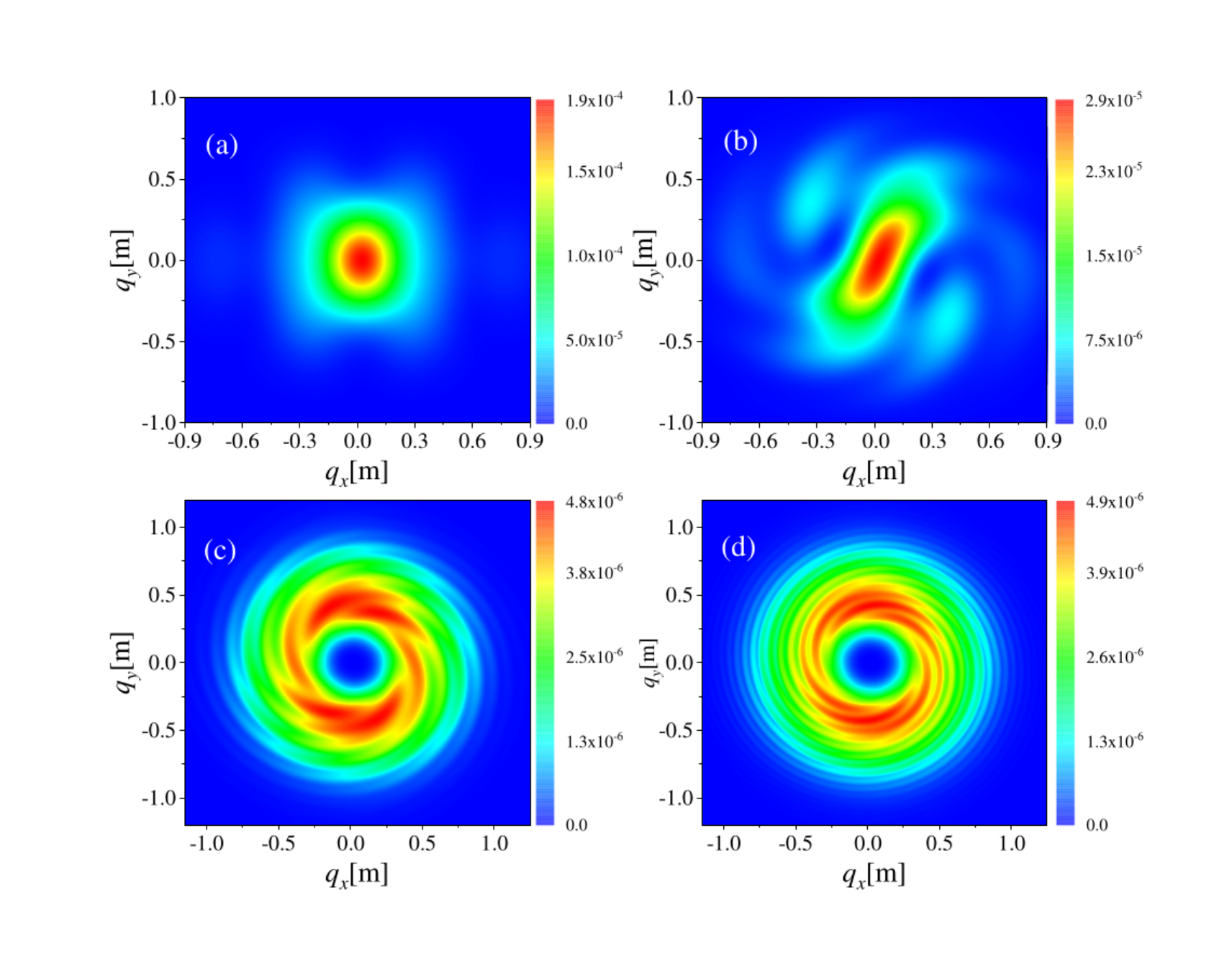}
\end{center}
\setlength{\abovecaptionskip}{-1.6cm}
\caption{(color online). Momentum spectra of created particles in the polarization plane (where $q_z=0$) for $N=1$ with different time delay parameters. From (a) to (d), the corresponding time delays are $T=0, \tau, 4\tau, 8\tau$, respectively. Other electric field parameters are the same as in Fig.~\ref{2}.}
\label{4}
\end{figure}

It is known that the cycles in pulse is relatively large in previous study \cite{Li:2017qwd,Li:2019hzi}, but here we consider the effect of time delay on pair production for small cycles. When $N=1$, the influences of different $T$ on the momentum spectra are displayed in Fig.~\ref{4}. For $T=0$, the phenomenon is similar to the cases of $N=2$ and $N=4$. But with increasing time delay, we discover some important phenomena. For $T=\tau$, the elliptic momentum distribution in Fig.~\ref{4}(a) is gradually distorted and elongated, at the same time, a pronounced interference phenomenon can be observed, see Fig.~\ref{4}(b). Moreover, the maximum value of momentum spectrum in Fig.~\ref{4}(b) is smaller than that in Fig.~\ref{4}(a). These results will be qualitatively interpreted by the semiclassical picture in the following paragraph.

Importantly, as time delay increases to $T=4\tau$ and $T=8\tau$, there exist still obvious spiral patterns consisting of eight spiral arms in the momentum spectra, as shown in Figs.~\ref{4}(c) and (d). It means that the Eq. (\ref{D4}) is also approximately applicable in the case of $N=1$. Moreover, compared to the cases of $T=\tau$ and $T=0$, one can see that the the maximum values of momentum spectra in Figs.~\ref{4}(c) and (d) are smaller than those in Figs.~\ref{4}(a) and (b). On the other hand, compared to the cases of $N=2$ and $N=4$, we found that the spiral patterns shrink significantly in the direction of small momentum. Because we know from electric field Eq. (\ref{model}) that when $N$ reduces, the corresponding time delay and the range of effective time action on the pair production also decrease, which leads to a smaller distribution of spirals.

\begin{figure}[H]
\centering
\subfigure{\includegraphics[width=6.02cm]{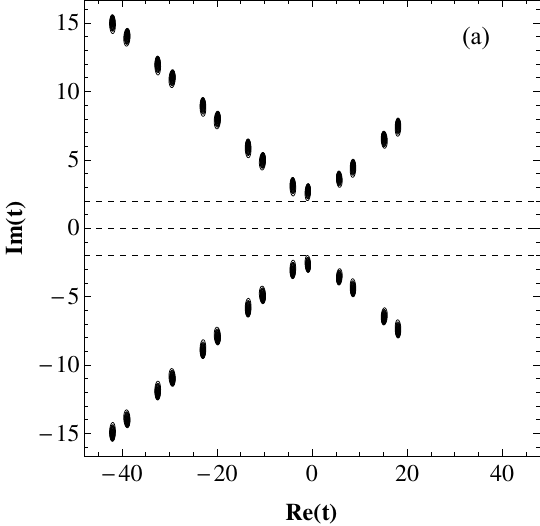}}
\hspace{4.6mm}
\subfigure{\includegraphics[width=6.02cm]{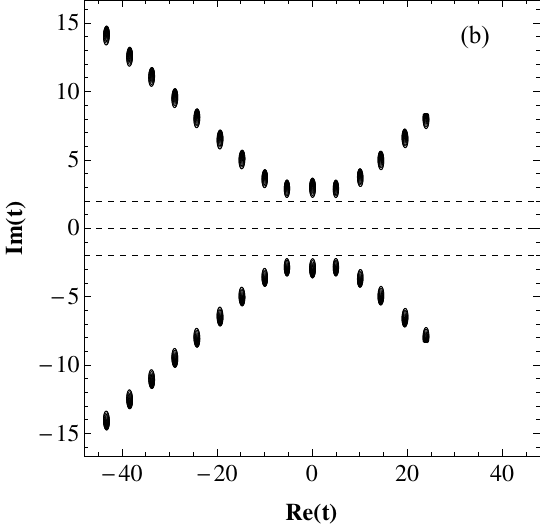}}
\hspace{4.6mm}
\subfigure{\includegraphics[width=6.02cm]{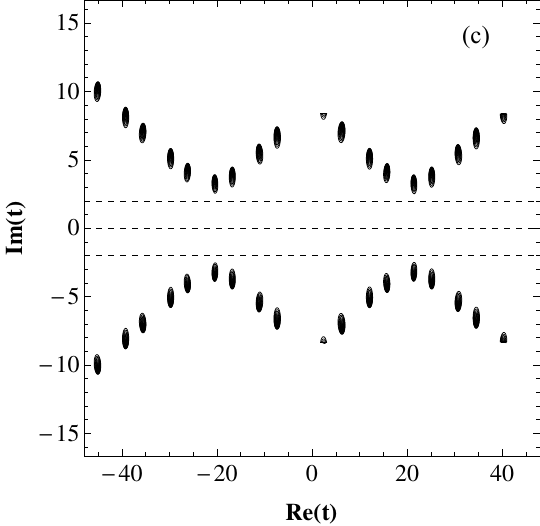}}
\hspace{4.6mm}
\subfigure{\includegraphics[width=6.02cm]{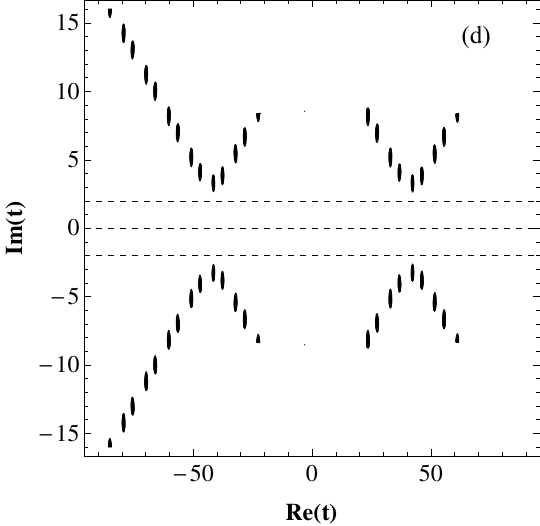}}
\setlength{\belowcaptionskip}{-0.3cm}
\caption{(color online). Contour plots of $|\Omega(\mathbf{q},t)|^2$ in the complex $t$ plane, showing the turning point distribution where $\Omega(\mathbf{q},t)=0$. These plots are for the cycle $N=1$, and other field parameters are the same as in Fig.~\ref{4}. From (a) to (d), the corresponding time delays are $T=0, \tau, 4\tau, 8\tau$, respectively, and the corresponding maximums of the momentum spectra are located in ($q_{x}=0.03, q_{y}=0.91$), ($q_{x}=0.03, q_{y}=0.02$), ($q_{x}=-0.14, q_{y}=-0.37$), ($q_{x}=-0.01, q_{y}=-0.43$), respectively. The three dashed lines are for eyes guidance.}
\label{5}
\end{figure}

It is known that the interference effects of momentum spectrum and the number density of created particles are associated with the location of turning points in the complex $t$ plane \cite{landau,heading, Brezin1970,Dumlu1011,Strobel2015}. Specifically, the number density depends on the turning points nearest to the real $t$ axis also called the dominant turning points, while the interference effects are dominated by the distances between the dominant turning points along the real $t$ axis direction. The turning points structures corresponding to the maximum momentum distribution for different $T$ in Fig.~\ref{4} are shown in Fig.~\ref{5}.
One can see that for $T=0$, there is one pair of dominant turning points, see Fig.~\ref{5}(a), while for $T=\tau$, we can observe three pairs of dominant turning points that they have almost equidistant along the real $t$ axis, see Fig.~\ref{5}(b). It is well known that the closer the distance between the dominant turning points are along the real axis, the stronger the interference effect of the momentum spectrum. Therefore, there is an obvious interference pattern in the momentum spectrum shown in Fig.~\ref{4}(b). Moreover, it is found that the dominant turning points in Fig.~\ref{5}(a) are closer to the real $t$ axis than those in Fig.~\ref{5}(b). And we known that the closer the dominant turning points is to the real axis, the greater the number density of created particles. Thus $n((q_{x}=0.03, q_{y}=0.91), t\rightarrow\infty)=1.91\times10^{-4}$ in Fig.~\ref{4}(a) is large than $n((q_{x}=0.03, q_{y}=0.02), t\rightarrow\infty)=2.87\times10^{-5}$ in Fig.~\ref{4}(b).

As time delay increases to $T=4\tau$ and $T=8\tau$, the distributions of turning points become more complicated as displayed in Figs.~\ref{5}(c) and (d). It is found that there are four pairs of dominant turning points and the distributions present an obvious periodic structure. Note that the four pairs of turning points are obtained by contributing two pairs per period. This means that the turning points distributions exist an interference within each period (second order interference) in addition to the interference between the two periods (first order interference). We think that the periodicity of turning points may be primarily related to the generation of spirals in the momentum spectra, while the combined effect of two orders interference may be mainly associated with the interference between the spiral arms. Therefore, one can see from Figs.~\ref{5}(c) and (d) that the distributions of turning points show a remarkable periodic structure, which leads the generation of the spirals in Figs.~\ref{4}(c) and (d).

Moreover, from Fig.~\ref{5}(c), it is found that the distance along the real $t$ axis direction between the dominant turning points of two periods is $\Delta\mathrm{Re}(t)\approx40$, while the corresponding distance in Fig.~\ref{5}(d) is $\Delta\mathrm{Re}(t)\approx85$. Meanwhile, the distance along the real $t$ axis direction between the two pairs of dominant turning points each period in Fig.~\ref{5}(c) is $\Delta\mathrm{Re}(t)\approx4$, while the corresponding distance in Fig.~\ref{5}(d) is $\Delta\mathrm{Re}(t)\approx5$. Therefore, the total interference of the turning points distribution in Fig.~\ref{5}(c) is stronger than that in Fig.~\ref{5}(d). It demonstrates that the interference effect between the spiral arms in Fig.~\ref{4}(c) is stronger than that in Fig.~\ref{4}(d). Besides, the turning points characteristics in Figs.~\ref{5}(c) and (d) also reflect the fact that the time delay between the two fields in the case of Fig.~\ref{5}(c) is smaller than that of Fig.~\ref{5}(d), which is consistent with the information reflected by our electric field Eq. (\ref{model}). On the other hand, compared to the case of $T=\tau$, one can see that the dominant turning points in Figs.~\ref{5}(c) and (d) are farther from the real $t$ axis than those in Fig.~\ref{5}(b), but the dominant turning points in Figs.~\ref{5}(c) and (d) have almost the same distances from the real axis. It suggests that $n((q_{x}=0.03, q_{y}=0.02), t\rightarrow\infty)=2.87\times10^{-5}$ in Fig.~\ref{4}(b) is large than $n((q_{x}=-0.14, q_{y}=-0.37), t\rightarrow\infty)=4.8\times10^{-6}$ in Fig.~\ref{4}(c) and $n((q_{x}=-0.01, q_{y}=-0.43), t\rightarrow\infty)=4.9\times10^{-6}$ in Fig.~\ref{4}(d), while $n((q_{x}=-0.14, q_{y}=-0.37), t\rightarrow\infty)=4.8\times10^{-6}$ in Fig.~\ref{4}(c) and $n((q_{x}=-0.01, q_{y}=-0.43), t\rightarrow\infty)=4.9\times10^{-6}$ in Fig.~\ref{4}(d) are almost equal.

\subsection{$N=0.8$ and $N=0.5$}

\begin{figure}[H]
\begin{center}
\includegraphics[width=\textwidth]{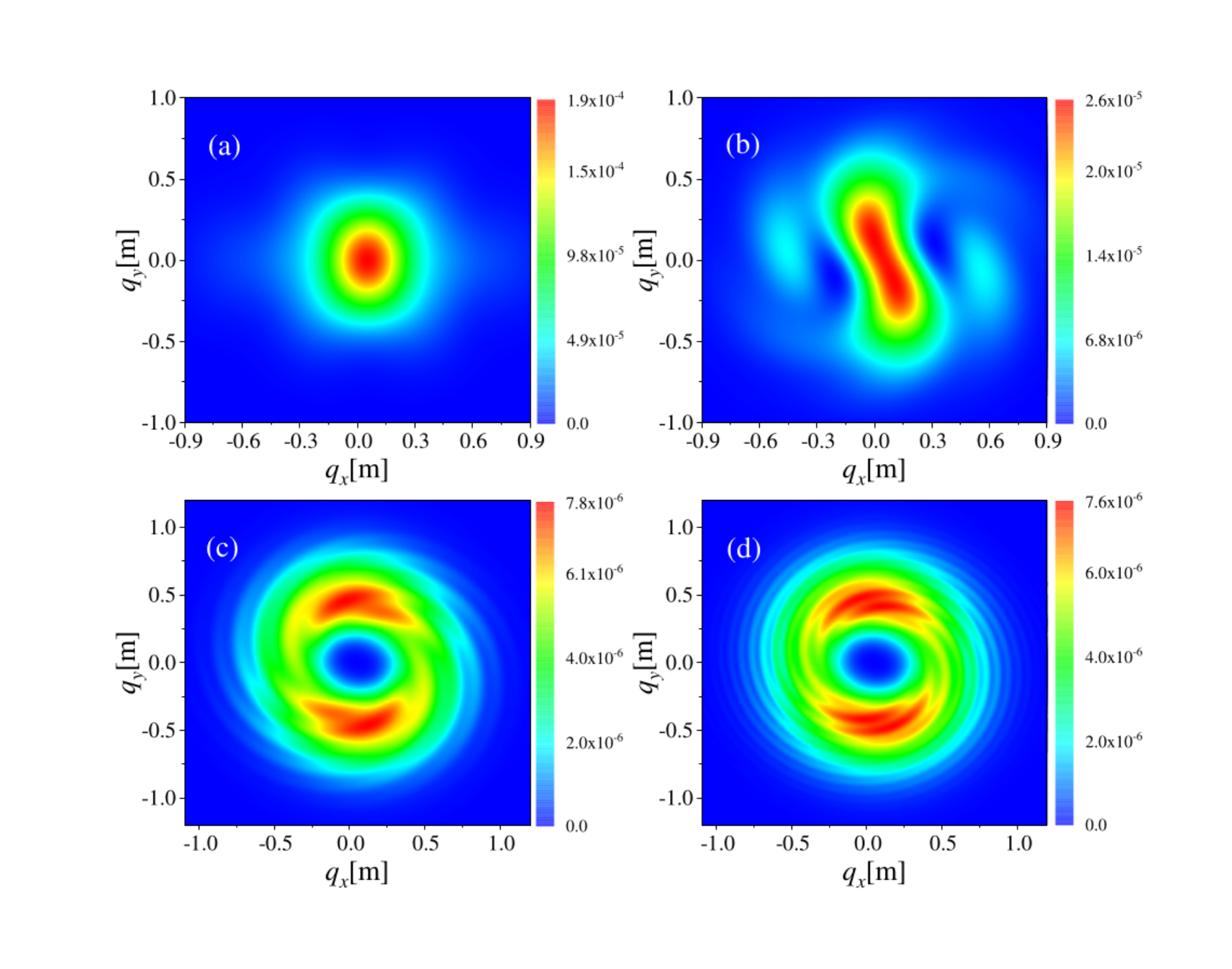}
\end{center}
\setlength{\abovecaptionskip}{-1.6cm}
\caption{(color online). Momentum spectra of created particles in the polarization plane (where $q_z=0$) for $N=0.8$ with different time delay parameters. From (a) to (d), the corresponding time delays are $T=0, \tau, 4\tau, 8\tau$, respectively. Other electric field parameters are the same as in Fig.~\ref{2}.}
\label{6}
\end{figure}

When the cycle decreases to $N=0.8$, we show the effects of $T$ on the momentum spectra in Fig.~\ref{6}. The results are almost similar to the case of $N=1$, except that for $T=4\tau$ and $T=8\tau$, the momentum spirals are less pronounced than those in the case of $N=1$. However, for $T=8\tau$, a spiral structure composed of eight spiral arms can still be generated in the momentum spectrum. It indicates that the appropriate time delay under the subcycle also causes the generation of spirals in the momentum spectrum, which provides a new reference for the possible experimental observation about the multiphoton pair production in future. Based on this finding, we further explore whether there is still a spiral when the cycle decreases to $N=0.5$?

\begin{figure}[H]
\begin{center}
\includegraphics[width=\textwidth]{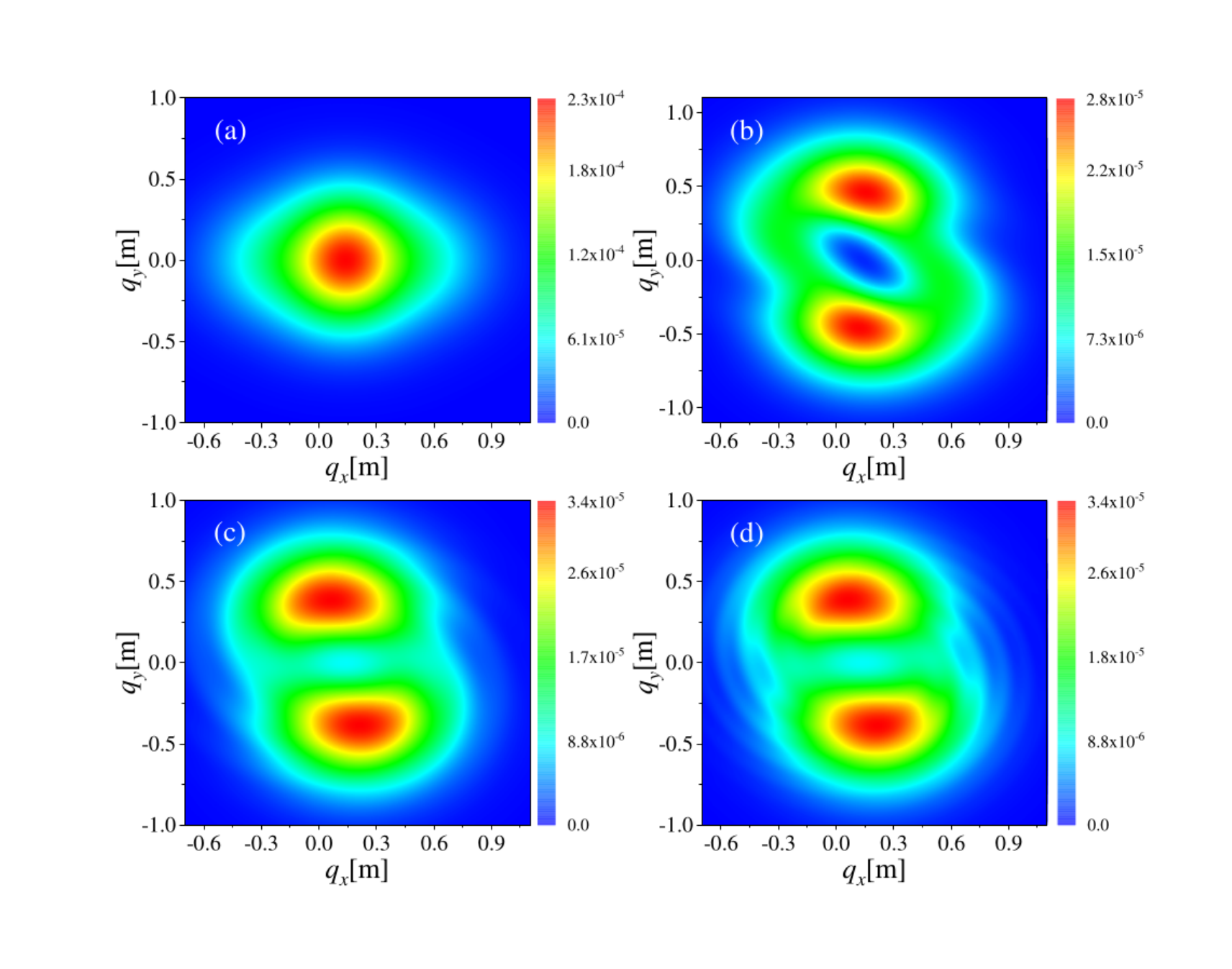}
\end{center}
\setlength{\abovecaptionskip}{-1.6cm}
\caption{(color online). Momentum spectra of created particles in the polarization plane (where $q_z=0$) for $N=0.5$ with different time delay parameters. From (a) to (d), the corresponding time delays are $T=0, \tau, 4\tau, 8\tau$, respectively. Other electric field parameters are the same as in Fig.~\ref{2}.}
\label{7}
\end{figure}

When the cycle decreases further to $N=0.5$, the effects of $T$ on the momentum spectra are shown in Fig.~\ref{7}. In Fig.~\ref{7}(a), we find a similar phenomenon as in Fig.~\ref{6}(a), i.e., the center of momentum spectra for created particles in the polarization plane is not located exactly at ($q_{x}=0, q_{y}=0$), but is shifted little. The reason is that when external field is turned off, a small value of vector potential ${\mathbf A}(t\rightarrow\infty)\neq0$ is possible for the very short subcycle pulse. Therefore, the symmetry point has a little deviation from ($q_{x}=0, q_{y}=0$). The phenomenon is similar to the cases in Fig.$1$ of the Ref. \cite{ Li:2021vjf} and in Fig.$2$ of the Ref. \cite{Hebenstreit:2009km}. In addition, from Fig.~\ref{7}, we found some new phenomena compared to the case of $N=0.8$. As can be seen in Fig.~\ref{7}(a), for $T=0$, there is one of the strongest momentum distribution that is near the center, while as time delay increases, the momentum distribution along the $q_{y}$ direction shifts rapidly toward the large momentum direction, at the same time, the strongest momentum distribution is split into two parts that are far from the center, see Figs.~\ref{7}(b), (c) and (d). The reason is that $e^{-}e^{+}$ pairs are mainly generated at the two maximums of the electric field, where $T=\pm G\tau$, $G=1,4,8$. Especially in Figs.~\ref{7}(c) and (d), we found that the range of momentum distribution along the $q_{x}$ direction expands and a weak interference appears. The interference can be interpreted as interference effect of particles created by large peaks of the two counter-rotating fields.

Importantly, compared to the case of $N=0.8$, it is found that even if the time delay increases to $T=8\tau$, there is still no pronounced spiral structure in the momentum spectrum. It indicates that spiral is very sensitive to the number of cycles, i.e., even if the time delay is large, the obvious spirals still cannot be generated if the cycle is very small. Meanwhile, from Fig.~\ref{7}, we found that the time delay mainly affects the momentum separation in the $q_{y}$ direction, and one can see from Fig.~\ref{6} that the number of cycles seems to primarily dominate the momentum distribution in the $q_{x}$ direction, while the combined effect of time delay and the number of cycles affects the generation of the spiral structure. Furthermore, in Fig.~\ref{7}, one can see a simple distribution, which has a no pronounced spiral structure in the momentum spectrum. This may be understood qualitatively from the the evolution curves in Figs.~\ref{1}(a) and (b). It is found that the the curves are not widely and uniformly distributed, especially, the curve in Fig.~\ref{1}(a) has no self crossing for the single field. This leads to a less obvious spiral structure in the corresponding momentum spectrum.

\section{\textcolor[rgb]{1.00,0.00,0.00}{Spirals} for fields with relative phase $\Delta\phi=\pi/2$}\label{Momentum2}

In this section, the influence of time delay with different cycles in pulse on the momentum spirals in two counter-rotating fields with relative phase $\Delta\phi=\phi_2-\phi_1=\pi/2$ are investigated. Note that since the results in the cases of $N=4$ and $N=2$ are almost similar to those for $\Delta\phi=0$, except that all patterns are rotated $\Delta\phi/2=\pi/4$ counterclockwise, so we do not show the results here. In the following, we are focusing on the study of the cases of $N=1$ and $N=0.5$.

\begin{figure}[H]
\begin{center}
\includegraphics[width=\textwidth]{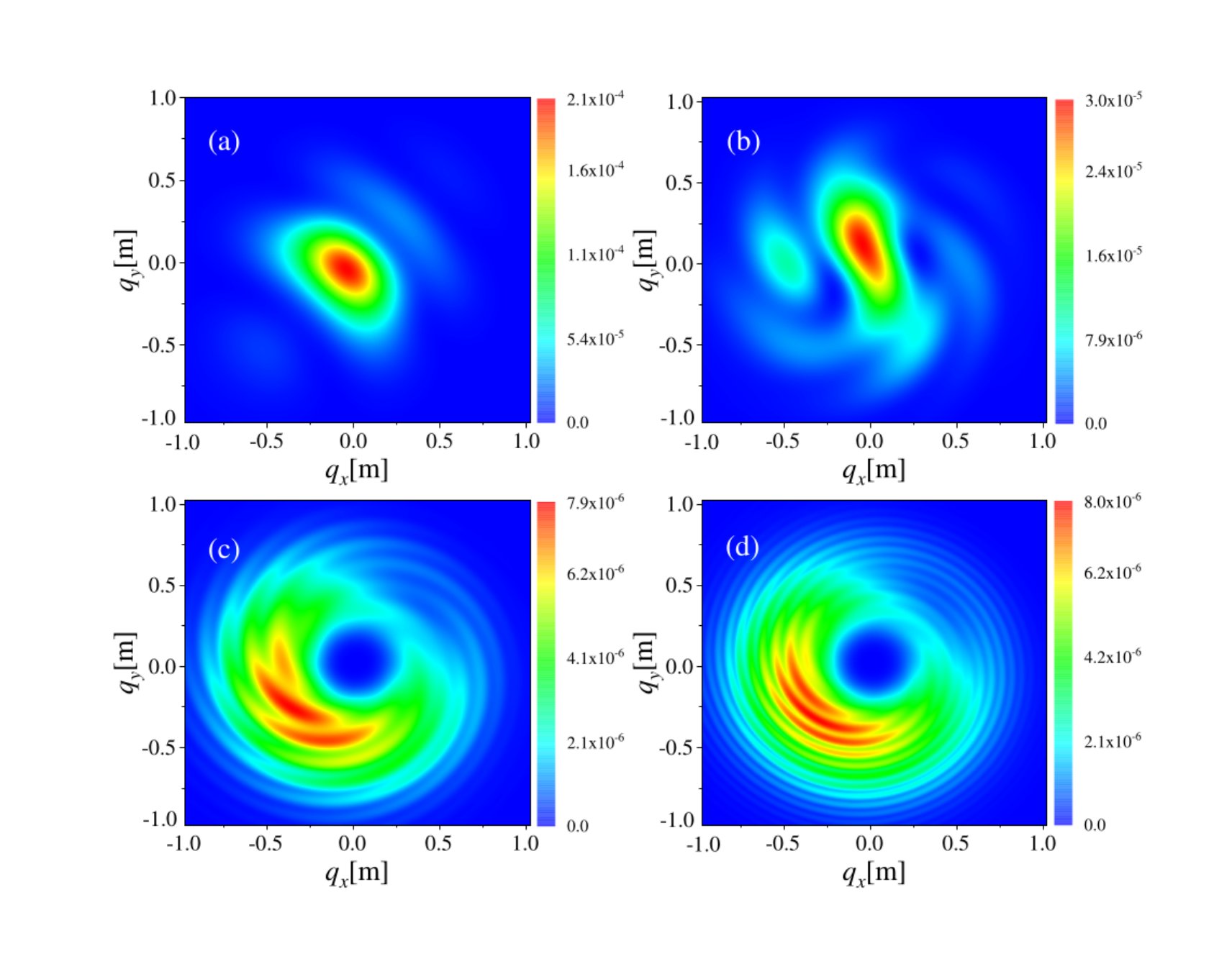}
\end{center}
\setlength{\abovecaptionskip}{-1.6cm}
\caption{(color online). Momentum spectra of created particles in the polarization plane (where $q_z=0$) for $N=1$ with different time delay parameters. The field parameters are the same as in Fig.~\ref{4} except $\phi_{2}=\pi/2$.}
\label{8}
\end{figure}

When $N=1$, the effects of $T$ on the momentum spectra are displayed in Fig.~\ref{8} where the remarkable difference could be observed. Firstly, for $T=0$, the axisymmetry of the momentum spectrum in the $q_{x}$ and $q_{y}$ directions is severely destroyed, but since $\phi_{2}=\pi/2$, the polarized axis are rotated $\Delta\phi/2=\pi/4$ counterclockwise to produce a new coordinates as ($q'_{x}$, $q'_{y}$), the symmetry in the $q'_{y}$ direction still exists while the symmetry in the $q'_{x}$ direction is broken, as shown in Fig.~\ref{8}(a). The reason is that under the new coordinates, as time reversal, the time $t$ and the momentums $q'_{x}$ and $q'_{y}$ change sign, the sign (odd/even property) of $\Omega(\mathbf{q'},t)=\sqrt{m^{2}+(q'_{x}-eA'_{x}(t))^{2}+(q'_{y}-eA'_{y}(t))^{2}}$ can still remain invariant only in the $q'_{y}$ direction, while its invariant is violated in the $q'_{x}$ direction. Therefore, the momentum spectrum presents an axisymmetry only in the $q'_{y}$ direction. Secondly, as time delay increases to $T=4\tau$ and $T=8\tau$, the rotational symmetry of spiral is also severely broken, since that the spiral pattern is mainly distributed in the third quadrant, see Figs.~\ref{8}(c) and (d). This phenomenon can be understood based on the knowledge of turning points. We know that the turning points structure is related to the solution of $\Omega(\mathbf{q'},t)=0$. For $T=4\tau$ and $T=8\tau$, when the polarized axis are rotated $\pi/4$ counterclockwise, the $\Omega(\mathbf{q'},t)$ can be eventually rewritten as $\Omega(\mathbf{q'},t)\approx \sqrt{m^{2}+f_1^2+f_2^2+{q'_{x}}^{2}+2\alpha q'_{x}}$ near the $q'_{y}=0$, here $\alpha \sim f_{1}(t)+f_{2}(t)$. Since that $\alpha>0$ thus it is easier to satisfy the equation $\Omega(\mathbf{q'},t)=0$ with $q'_{x}<0$, which means that the turning points are closer to the real $t$ axis in the region of $q'_{x}<0$ that leads to the dominant momentum locating the third quadrant. Moreover, we found that the number of corresponding spiral arms is significantly decreased and the shape of arms becomes slender.

\begin{figure}[H]
\begin{center}
\includegraphics[width=\textwidth]{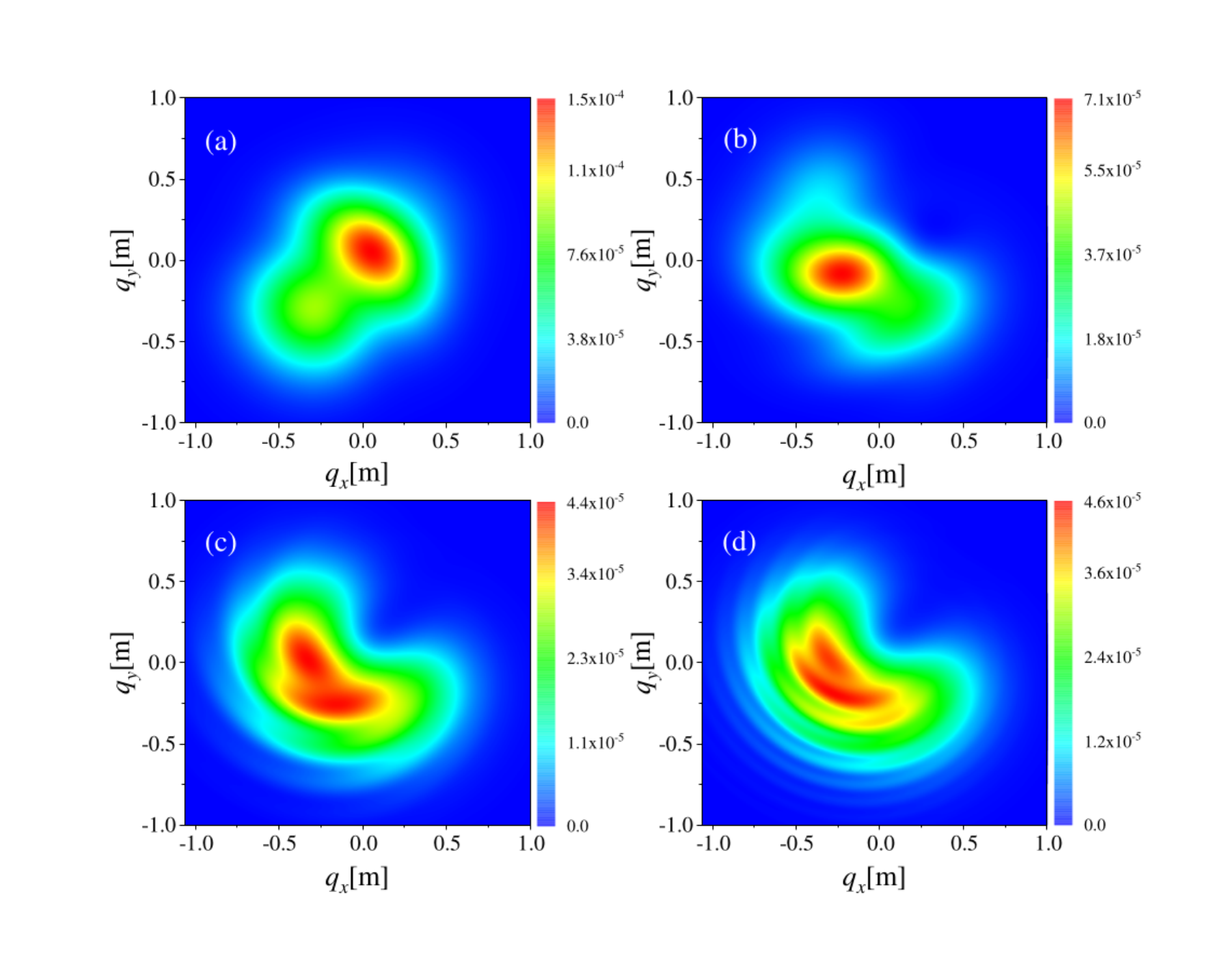}
\end{center}
\setlength{\abovecaptionskip}{-1.6cm}
\caption{(color online). Momentum spectra of created particles in the polarization plane (where $q_z=0$) for $N=0.5$ with different time delay parameters. The field parameters are the same as in Fig.~\ref{7} except $\phi_{2}=\pi/2$.}
\label{9}
\end{figure}

When the cycle decreases to $N=0.5$, the influences of $T$ on the momentum spectra are shown in Fig.~\ref{9}. Compared to the case of $\phi_{2}=0$ in Fig.~\ref{7}, we discover some interesting phenomena except that the axisymmetry of momentum distribution in Fig.~\ref{9}(a) is severely destroyed. With the increase of time delay, the maximum momentum distribution that was originally split into two parts in Fig.~\ref{7} is merged into one part, and the range of distribution is significantly shrunken, see Figs.~\ref{9}(b), (c) and (d). This phenomenon is due to the fact that when $\phi_{2}=\pi/2$, there is always only one maximum field strength in the electric field Eq. (\ref{model}) as the time delay increases, and $e^{-}e^{+}$ pairs are mainly created at the maximum of the electric field. Therefore, there exists only one maximum momentum distribution in the corresponding momentum spectra. Interestingly, for $T=8\tau$, we found that spiral structure is still generated in the momentum spectrum. However, when $\phi_{2}=0$, there are no spirals in this case. It indicates that the introduction of carrier phase can lead to the generation of spiral in the momentum spectrum even if the cycle is very small.

Of course, another interesting point is that we found the range of critical polarization values for the appearance of spirals in the momentum spectra for different cycles with $\Delta\phi=0$ and $\Delta\phi=\pi/2$ by numerical calculations. The results are shown in Table~\ref{Table 2}, note that where we only consider $T=8\tau$ for each cycle. It is found that in two cases of relative phases, when the cycle decrease from $N=2$ to $N=0.5$, the polarization range for the transition from the appearance to the disappearance of spirals is gradually decreasing. Moreover, for $N=0.5$, there is always no spiral in the momentum spectrum when $\Delta\phi=0$, while when $\Delta\phi=\pi/2$, momentum spectrum exists spiral, and the polarization range of the spiral transition is $0.4\thicksim0.3$.

\begin{table}[H]
\caption{Critical polarization range of the transition of spirals appearance/disappearance for various cycles with $\Delta\phi=0$ and $\Delta\phi=\pi/2$, respectively, when $T=8\tau$ is given. Note that the $--$ denotes the absence of spiral under the studied parameters.}
\centering
\begin{ruledtabular}
\begin{tabular}{cccc}
number of cycles   &$N=2$   &$N=1$  &$N=0.5$ \\
\hline
$|\delta_{1,2}|$ ($\Delta\phi=0$)       &$0.6\thicksim0.5$   &$0.5\thicksim0.4$  &$--$  \\
$|\delta_{1,2}|$ ($\Delta\phi=\pi/2$)   &$0.6\thicksim0.5$   &$0.5\thicksim0.4$  &$0.4\thicksim0.3$  \\
\end{tabular}
\end{ruledtabular}
\label{Table 2}
\end{table}

\begin{figure}[H]
\begin{center}
\includegraphics[width=\textwidth]{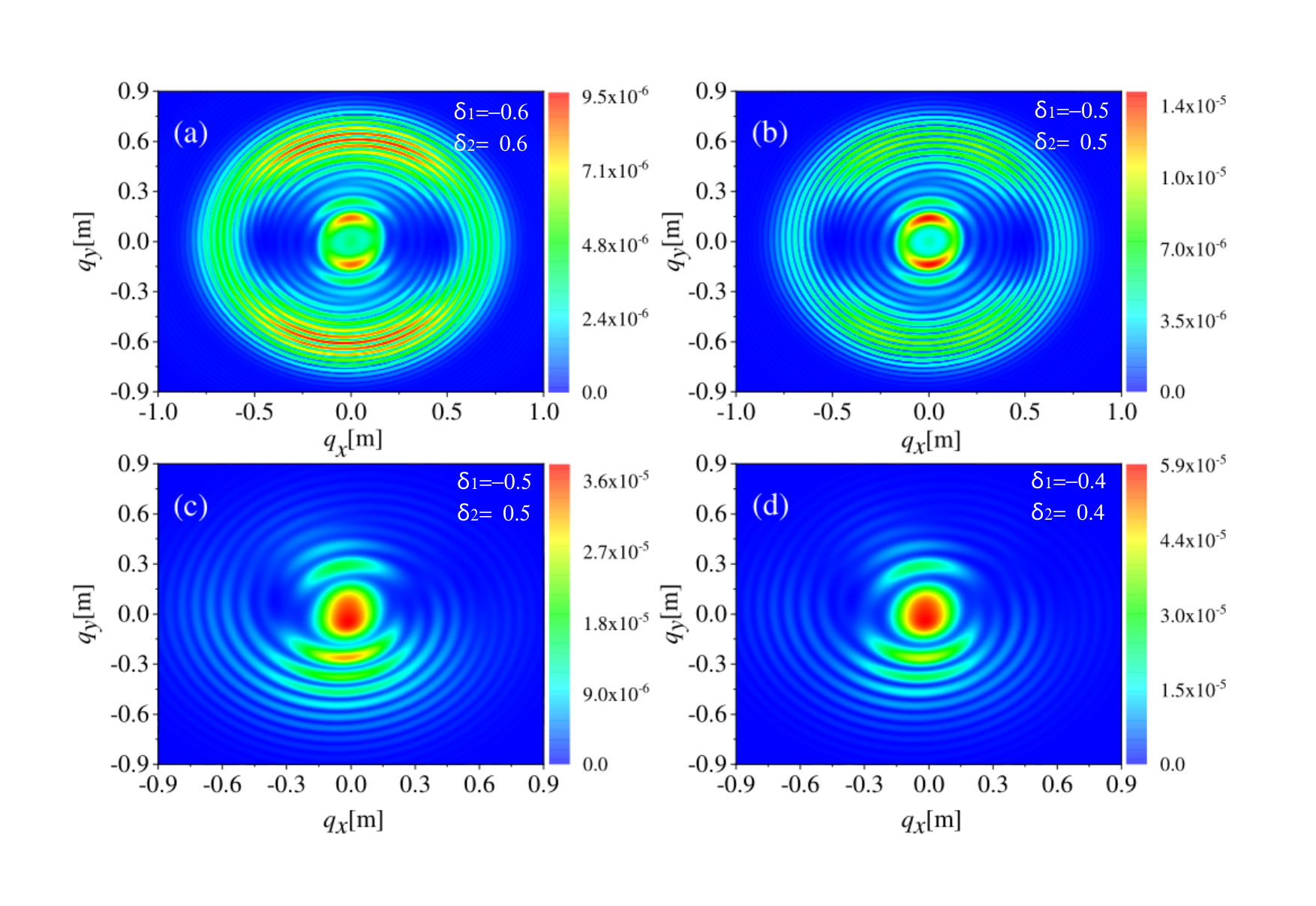}
\end{center}
\setlength{\abovecaptionskip}{-1.6cm}
\caption{(color online). Momentum spectra of created particles in the polarization plane (where $q_z=0$) for various polarization value with different $N$ and $\Delta\phi$. Where the time delay is set as $T=8\tau$, (a) and (b) correspond to the case of $N=2$ and $\Delta\phi=0$, (c) and (d) correspond to the case of $N=1$ and $\Delta\phi=\pi/2$.}
\label{10}
\end{figure}

In order to observe the change in polarization values during the spiral transition, we select some examples from Table~\ref{Table 2} for the study, as shown in Fig.~\ref{10}. One can see from Figs.~\ref{10}(a) and (b) that for $N=2$ with $\Delta\phi=0$, there are still spiral in the momentum spectrum as $|\delta_{1,2}|$ decreases to $0.6$, while when $|\delta_{1,2}|$ further decreases to $0.5$, the spiral gradually disappear and the momentum distribution shrinks primarily toward small momentum direction. From Figs.~\ref{10}(c) and (d), it is found that for $N=1$ with $\Delta\phi=\pi/2$, the momentum spectrum still exists spiral as $|\delta_{1,2}|$ reduces to $0.5$, while when $|\delta_{1,2}|$ further reduces to $0.4$, the spiral gradually disappear. These results show that we can observe spiral patterns not only in the two counter-rotating CP fields ($|\delta_{1,2}|=1$), but also in the two counter-rotating elliptical polarization fields ($|\delta_{1,2}|\thickapprox0.5$). It greatly reduces the polarization of the field to observe spirals effectively.

It should be noted that $N=4$ is a special case, in which there are no the spiral transition, but a variation of spiral split. The details are described as follows, we can observe spiral patterns at all polarizations, but at $|\delta_{1,2}|=1$, the spiral structure is relatively uniform and is a globe, while at $|\delta_{1,2}|\in[0.9,0.5]$, it is split into two parts, and at $|\delta_{1,2}|\in[0.4,0]$, it is further split into four parts. Here we do not display the results.

Based on the effect of field polarization on spiral formation and change mentioned above, it is reminding that by adjusting the polarization, we can control not only the appearance or disappearance of spiral pattern, but also the location of spiral presence.

\section{Number density}\label{density}

In this section, the effects of time delay and cycle in pulse on the number density of created particles in the case of relative phase $\Delta\phi=0$ and $\Delta\phi=\pi/2$ are investigated, respectively. Note that for comparison between the results, the following studies are performed under the same laser field energy \cite{Nuriman:2015}, and we selected several different field parameters.

\begin{figure}[H]
\begin{center}
\includegraphics[width=0.9\textwidth]{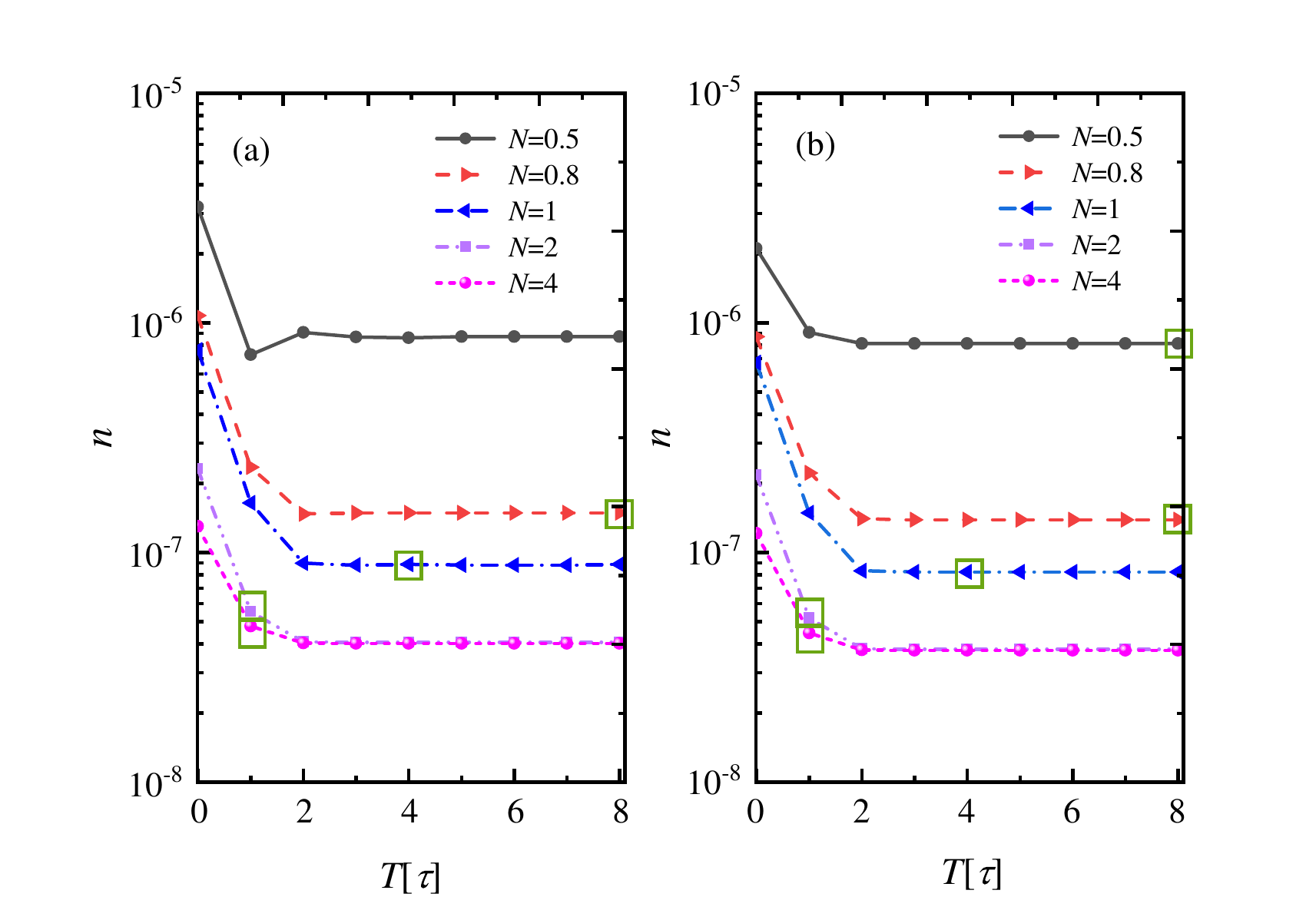}
\end{center}
\setlength{\abovecaptionskip}{-1.0cm}
\caption{(color online). Number density of created particles under the same laser field energy dependence on time delays for various cycles with different phase parameters $\Delta\phi=0$ for (a) and $\Delta\phi=\pi/2$ for (b), respectively. Other electric field parameters are the same as in Fig.~\ref{2}. Note that the green rectangle marks the minimal $T$ when the spiral is generated.}
\label{11}
\end{figure}

The number density dependence on $T$ for various $N$ is shown in Fig.~\ref{11}. In the case of $\Delta\phi=0$, one can see from Fig.~\ref{11}(a) that when $T$ is fixed, the number density does not change obviously at large $N$, while it is significantly enhanced about one order of magnitude at small $N$. When $N$ is fixed, the number density tends to be a constant at large $T$, while it is increased at least five times at small $T$. Combining the above we can conclude that either the small $T$ or $N$ is beneficial for $e^{-}e^{+}$ pair production. On the other hand, according to the markings of the green rectangle in the figure, we found that when $N$ is larger, instead, spirals start to be generated in the momentum spectrum at smaller $T$. In particular, when $N=0.8$ and $N=1$, the corresponding number densities only differ about $2$ times, but the minimal $T$ for spiral generation are significantly different. Specifically, when $N=0.8$, a spiral structure appears in the momentum spectrum at $T=8\tau$, while when $N=1$, it appears in the momentum spectrum at $T=4\tau$. It indicates that without losing much number density, we can obtain the spiral pattern at a smaller time delay by adjusting the above two parameters flexibly.

In the case of $\Delta\phi=\pi/2$, the results shown in Fig.~\ref{11}(b) are similar to those of Fig.~\ref{11}(a). The difference is that for $T=0$, the number density of $\Delta\phi=0$ is slightly larger than that of $\Delta\phi=\pi/2$ for $N=0.5$ and $N=0.8$. Moreover, for $N=0.5$, there is no spiral structure in the case of $\Delta\phi=0$, while in the case of $\Delta\phi=\pi/2$, momentum spectrum exists spiral pattern. This indicates that the introduction of the phase has little effect on the number density of the generated particles, while it mainly affects the momentum spirals.

\begin{figure}[H]
\begin{center}
\includegraphics[width=0.9\textwidth]{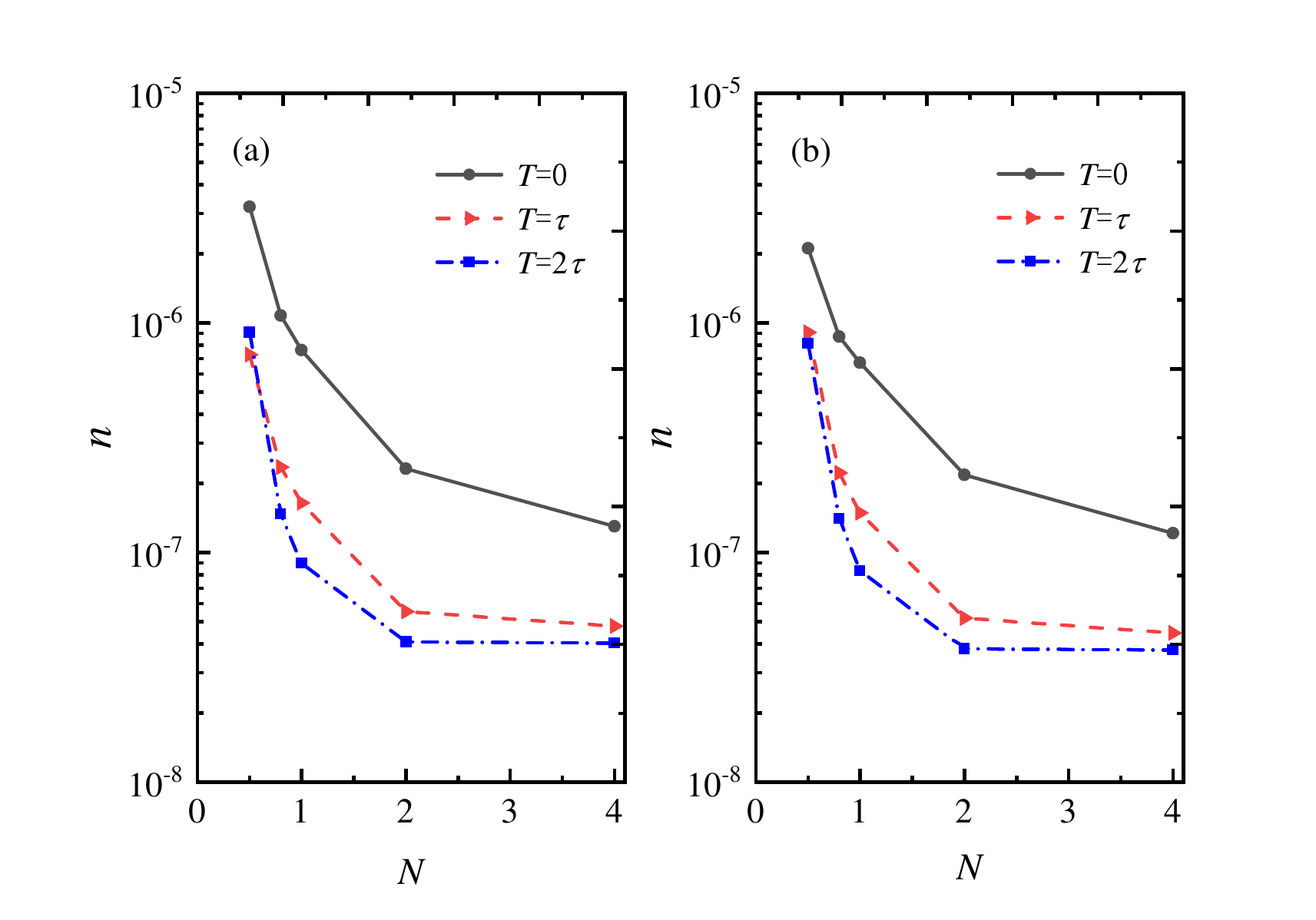}
\end{center}
\setlength{\abovecaptionskip}{-1.0cm}
\caption{(color online). Number density of created particles under the same laser field energy dependence on cycles in pulse for small time delay with different relative phase parameters. The electric field parameters are the same as in Fig.~\ref{11}.}
\label{12}
\end{figure}

To see more clearly how the number density of created particles varies with $N$ for small $T$, we display Fig.~\ref{12}. It is found that in the cases of $\Delta\phi=0$ and $\Delta\phi=\pi/2$, when $T$ is fixed, the corresponding number density is enhanced about one order of magnitude with the decrease of $N$, while when $N$ is fixed, it is increased by few times with reducing $T$. These results indicate that the number density is more sensitive to the number of cycles in pulse.

\section{CONCLUSION AND OUTLOOK}\label{conclusion}

In summary, we revisit the spirals in multiphoton pair creation by two counter-rotating fields with time delay for different cycles using the DHW formalism. The focus is on considering two case of the relative carrier envelope phase as $0$ and $\pi/2$, and the effects of different time delays and cycles on the number density are further examined. Meanwhile, some typical spiral structures are semiquantitatively analyzed by employing the WKB-like approximation method. Moreover, we provide some qualitative understandings to some results obtained by corresponding turning points structure.

For the momentum spiral, it is sensitive to time delays and cycles in pulse. compared to previous studies \cite{Li:2017qwd,Li:2019hzi}, we found some interesting new results. With the increase of either time delay or cycle number, the spiral arms become thinner and longer, meanwhile, the number of spiral arms is significantly increased. Importantly, for a small cycle $N=0.8$, the momentum spectrum still exists an obvious spiral pattern. Moreover, the carrier phase plays an crucial role in multiphoton pair production, which destroys not only the axisymmetry of momentum spectrum, but also the rotational symmetry of momentum spiral. And the number of spiral arms is decreased due to the introduction of carrier phase. More importantly, for $\Delta\phi=\pi/2$, there still exists spiral pattern in the momentum spectrum when the cycle decreased to $N=0.5$. On the other hand, we also found the range of critical polarization values for the spirals appearance corresponding to the different cycle number with relative carrier envelope phase as $0$ and $\pi/2$, respectively. Based on the studied results, it is applicable to regard the momentum signatures as a new probing to the laser field information. For example, by breaking the symmetry, we can probe the information about relative phase. By the thinning shape and the number of spiral arms, one can detect the information of cycles in pulse. By the presence or absence of spirals in the momentum spectrum, we can probe the information of time delay.

For the number density of created particles, it is insensitive to relative phase, but sensitive to time delay and cycle number. We found that either small time delay or cycle increases the number density significantly. Specifically, the number density is increased at least five times at small time delay, and it is enhanced about one order of magnitude at small cycle. While for either large time delay or pulse cycle, the number density tends to be a constant. Interestingly, it is found that without losing much number density, we can obtain the spiral pattern at a smaller time delay by adjusting the above two parameters flexibly. This is important since it may provide a possibility of broader parameter ranges for realizing the spirals in multiphoton pair production.

These results indicate that time delay, the number of cycles in pulse and carrier envelope phase play an extremely important role in spirals of multiphoton pair creation by two counter-rotating fields. While we have only investigated two typical cases of the carrier phase, we believe that the results have exhibited many important features about the spirals of multiphoton pair production.

\begin{acknowledgments}
\noindent
We are thankful to the anonymous referee for helpful suggestions to improve the manuscript. This work was supported by the National Natural Science Foundation of China (NSFC) under Grant No.\ 11935008 and No.\ 11875007. The computation was carried out at the HSCC of the Beijing Normal University.
\end{acknowledgments}

\end{document}